\documentstyle[preprint,aps,eqsecnum]{revtex}

\newcommand{\tr}{{\rm Tr}}
\newcommand{\del}{\partial}
\newcommand{\deldel}{\tensor{\del}}
\newcommand{\ps}{{p \hspace{-5pt}/}}
\newcommand{\qs}{{q \hspace{-5.5pt}/}}

\newcommand{\rs}{{r \hspace{-7pt}/}}

\begin{document}
\setcounter{page}{1}

\draft

\title{
 Hadronic Light-by-light Scattering Contribution
  to Muon $ g - 2 $}

\author{
 M. Hayakawa$ ^1$ \footnotemark[1],
 T. Kinoshita$ ^2$ \footnotemark[2], 
 and A. I. Sanda$ ^1$ \footnotemark[3] }
\footnotetext[1]
{
 Electronic address: hayakawa@eken.phys.nagoya-u.ac.jp
}
\footnotetext[2]
{
 Electronic address: tk@hepth.cornell.edu
}
\footnotetext[3]
{
 Electronic address: sanda@eken.phys.nagoya-u.ac.jp
}

\address{
 $ ^1 $  Department of Physics, Nagoya University,
         Nagoya 464-01 Japan \\
 $ ^2 $  Newman Laboratory, Cornell University,
         Ithaca, New York 14853 USA }

\preprint{\begin{minipage}{4cm}
           DPNU-95-30 \\
           January 1996 \\
          \end{minipage}
         }

\date{January 21,\ 1996}

\maketitle

\begin{abstract}
 The hadronic light-by-light scattering
contribution to 
muon $g-2$ is examined based
on the low energy effective theories of QCD,
the Nambu-Jona-Lasinio model
and hidden local symmetry approach, 
supplemented by a general information
concerning the asymptotic behavior of QCD. 
 Our result is $- 52 \times 10^{-11}$ 
with an uncertainty of $\pm 18 \times 10^{-11}$, 
which includes our best estimate of model dependence.   
 This is within the expected measurement uncertainty 
of $40\times 10^{-11}$ 
in the forthcoming experiment
at Brookhaven National Laboratory.  
 Our result removes one of the main theoretical obstacles
in verifying the existence of 
the weak contribution to the muon $g-2$.
\end{abstract}


\narrowtext

\section{Introduction}
\label{Sec:intro}
%

A substantial improvement in the measurement
of the muon anomalous magnetic moment
$ a_\mu \equiv \frac{1}{2} \left( g_\mu - 2 \right) $ 
is planned
at the Brookhaven National Laboratory.
  The precision of the measurement
is expected to reach the level of \cite{Hughes}
\begin{equation}
 40 \times 10^{-11} .
 \label{expectederror}
\end{equation}
 This is about 20 times more accurate
than the best value available at present
\cite{CERN}
\begin{equation}
 a_\mu ({\rm exp}) = 1\ 165\ 923\ (8.5) \times 10^{-9} ,
\end{equation}
where the numerals in the parentheses represent the uncertainties
in the last digits of the measured value.

 Compared with the electron anomaly,
for which all contributions other than QED
are negligible,
 the muon anomaly is more sensitive
to shorter scales where the hadronic and weak interaction effects
are important.
 Also, provided that the standard model prediction
is known precisely,
the muon anomaly will be
a sensitive probe of physics  beyond the standard model.
 A typical standard model prediction is 
\cite{KM}
\begin{equation}
 a_\mu ({\rm th}) = 116\ 591\ 877 (176) \times 10^{-11}.
 \label{eq:prediction}
\end{equation}
 We note that the uncertainty in (\ref{eq:prediction})
is comparable
with the one-loop weak interaction correction \cite{weak}
\begin{equation}
 a_\mu({\rm weak}\mbox{-}1)=195\ (1) \times 10^{-11}.
\end{equation}
 Thus further improvement of the theoretical prediction is
necessary
in order to be able to confirm
the existence of the weak correction term in $a_\mu$.

 The uncertainty in (\ref{eq:prediction}) is dominated by 
the error associated with
the estimate of the 
strong interaction correction to
 $ a_\mu $.
 The bulk of this effect
is due to the hadronic vacuum polarization contribution,
which starts at $ {\cal O}(\alpha^2) $.
(See Fig. 1 of Ref. \cite{K-had} for the Feynman graphs 
which give 
this type of
contribution.) 
 Fortunately,
this contribution is calculable
without relying on 
our theoretical knowledge of strong interaction.
 The $ {\cal O}(\alpha^2) $
contribution to the $ a_\mu({\rm had.v.p.}) $ can be expressed 
in the form
\cite{Gourdin} 
\begin{equation}
 a_\mu({\rm had.v.p.}) \left|_{{\cal O }(\alpha^2)} \right.
  = \left(
      \frac{\alpha m_\mu}{3\pi}
    \right)^2
    \int_{4 m_\pi^2}^{\infty} ds
    \frac{R(s) K(s)}{s^2}
 \label{had-formula}
\end{equation}
by applying the dispersion relation
and the optical theorem.
 Here $ R(s) $ is the hadron production cross section
in $ e^+ e^- $ collisions
normalized to the lowest order formula for the $ \mu^+ \mu^- $
production cross section 
$ \sigma(e^+ e^- \rightarrow \mu^+ \mu^-) $
$ = 4\pi \alpha^2/3s $.
 The formula (\ref{had-formula}) enables us to reduce
the issue of our ignorance 
of strong interaction dynamics
to the experimental determination of $ R(s) $ \cite{had-theory}.
 The integral (\ref{had-formula})
has been evaluated by several groups \cite{K-had,groups,Eidelman}. 
For instance the estimate given in Ref.\cite{KM} is
\begin{equation}
 a_\mu({\rm had.v.p.})=7\ 068\ (59)\ (164)\times 10^{-11} ,
  \label{hvp}
\end{equation}
where the first and second errors are statistical and systematic,
respectively.
Works \cite{Eidelman,worstell}
which include more recent data are not too far off from (\ref{hvp}), 
although the evaluation of uncertainties
in the experimental data still varies considerably among authors.  
 Future measurements at VEPP-2M, DA$\Phi$NE,and BEPS
are expected to reduce these uncertainties to the level
of the upcoming experiment (\ref{expectederror})
\cite{worstell,Franzini}.

 On the other hand, 
the contribution of the hadronic light-by-light scattering diagram
shown in Fig. \ref{fig:light}
is potentially a source of more serious difficulty because 
it cannot be expressed in terms of 
experimentally accessible observables
and hence must be evaluated by purely theoretical consideration.  
 The purpose of this paper is to
report on our attempt to estimate
this hadronic light-by-light scattering contribution
to the muon anomaly.
 A summary of our preliminary results has been given in Ref. \cite{HKS}.
 We present the detailed analysis here.

 The paper is organized as follows.
 Sec. \ref{sec:review} starts with a survey of
previously reported results
on the hadronic light-by-light scattering contribution
to the muon $g - 2$.
 With the help of chiral perturbation theory and Nambu-Jona-Lasinio
(NJL) model,
we find that the relevant diagrams associated with
this contribution are the ones shown in Fig. \ref{fig:diagrams}
of Ref.\cite{deRafael}.
 We also give an outline of 
strategies to solve the problems we have encountered.
 The next three sections
are devoted to the treatment of the three types of diagrams;
 Sec. \ref{sec:loop} to the charged pseudoscalar loop
contribution of
Fig. \ref{fig:diagrams}(a),
 Sec. \ref{sec:pole} to the neutral pseudoscalar pole contribution
of Fig. \ref{fig:diagrams}(b),
 and Sec. \ref{sec:quark-loop} to the quark loop contribution
of Fig. \ref{fig:diagrams}(c).
 Sec. \ref{sec:summary} summarizes the present study
and compares it with the recent result
of Ref.\cite{Bijnens,Bijnens2} based
on the extended Nambu-Jona-Lasinio (ENJL) model 
and discuss its implications.

\section{Survey and Improvements}
\label{sec:review}

 This section begins with an overview of the previous studies
on the hadronic light-by-light scattering contribution to
the muon anomaly.
 We then point out a few problems associated with its evaluation,
and describe the procedure which we have adopted to solve them.

\subsection{Previous Studies}
\label{subsec:overview}

 The muon anomalous magnetic moment receives important contributions
from hadronic physics.
 Naive dimensional consideration suggests that the effect
of the physics
of the typical scale $ \Lambda $ higher than the muon mass 
$ m_\mu $
is suppressed by $ (m_\mu/\Lambda)^2 $. 
 This implies that
contributions to $ a_\mu $ from QCD
will be dominated by nonperturbative aspects of QCD.
 Thus we are confronted with a calculational difficulty;
the relevant hadronic contribution
to the light-by-light scattering amplitude 
may not be calculable from first principles 
in the current stage of development of QCD.

As the next best procedure, 
 we may appeal to the 
 chiral perturbation theory which attempts to describe
the low energy dynamics of QCD in terms of hadrons.
 Its leading behavior is given unambiguously by
the low energy theorems on the dynamics of pions (and kaons)
which are the Nambu-Goldstone bosons resulting from
spontaneous breakdown of chiral symmetry.
 The scalar QED calculation in Ref. \cite{K-had} corresponds
to the lowest-order evaluation in this context.
 Corrections to the lowest order results may be obtained by
adding higher order terms
of a power series expansion in momentum variables.

 For the calculation of the muon $g - 2$,
however, such a systematic chiral perturbation technique
runs into some problems.  
 Insertion of a vertex with high power of momentum
into Feynman diagrams for the muon anomaly,
as a correction to
the hadronic light-by-light scattering amplitude,
yields a divergent result.
 Thus we must resort to an alternative approach
which unfortunately is more model-dependent.  
 For instance, Ref. \cite{K-had} introduced the vector meson resonances.
 It should be noted that
the explicit incorporation of 
vector mesons allows one to compute higher order 
 counterterms  \cite{Ecker} in the chiral Lagrangian. The resulting
 $ {\cal O}(p^4) $ counter terms agree reasonably
well with experimental determination.
 A well-known example of the success in this direction
can be seen
in the description of pion's electromagnetic form factor 
$ F(q^2) $, where $ q $ is the photon momentum.
 There, the vector meson dominance (VMD) model works 
even for $ q $ as large as the mass of $ \rho $ meson,
$ M_\rho \simeq 760~{\rm MeV} $ .

 Now we shall return to our topic.
 From the point of view of chiral perturbation theory,
pions will contribute 
to $ a_\mu $ 
most significantly 
in the form of the diagrams shown
in Fig. \ref{fig:diagrams}(a) and (b).
 A priori we do not know the magnitudes of photon momenta
which are important for these contributions.
 For example one may attempt
to estimate the contribution of Fig. \ref{fig:diagrams}(a)
in the lowest-order of chiral expansion
which we will denote as  $ a_\mu(a,{\rm sQED})$.
 On the other hand we recognize that the VMD model
describes the $ \pi^+ \pi^- \gamma $ coupling well
for on-shell pions.
 Thus we are motivated to include the VMD model
explicitly in the $ \pi \pi \gamma $ coupling.
 A naive approach, which leads to 
 $ a_\mu(a, {\rm nVMD})$,
introduces vector meson
to replace a photon propagator as \cite{K-had} ;
\begin{equation}
 \frac{i}{q^2} \rightarrow
  \frac{i}{q^2} \frac{M_\rho^2}{M_\rho^2 - q^2}
  = \frac{i}{q^2} - \frac{i}{M_\rho^2 - q^2} \ .
 \label{eq:sub}
\end{equation}

 The numerical results obtained
by following these procedures were \cite{from-K-had}
\begin{eqnarray}
 a_\mu(a, {\rm sQED}) &=&
  \displaystyle{
   - 0.043\ 7\ (36) \times \left( \frac{\alpha}{\pi} \right)^3
  } 
   \nonumber \\
  &=&
  \displaystyle{
   - 54.76\ (46) \times 10^{-11}
  } ,
 \label{KNO-sQED}
\end{eqnarray}
and
\begin{eqnarray}
 a_\mu(a, {\rm nVMD}) &=&
  \displaystyle{
   - 0.012\ 5\ (19) \times \left( \frac{\alpha}{\pi} \right)^3
  } 
   \nonumber \\
  &=&
   \displaystyle{
    - 15.67\ (2.38) \times 10^{-11} 
   } ,
 \label{KNO-loop}
\end{eqnarray}
respectively. 
 We see a large modification when vector mesons are introduced.
A natural question arising from this observation is :

 ($Q_1$) Is the modification caused by the introduction
of VMD model real ?
 If it is, why it seems to conflict with our expectation based
on chiral perturbation theory that the
 vector meson effect is very small at low energies ?

 Next let us turn our attention to the diagram shown in
Fig. \ref{fig:diagrams}(b).
 It includes the  $ \pi^0 \gamma \gamma $ vertex
induced by the chiral anomaly.
 It is well-known that the effective interaction
\begin{equation}
 {\cal L} = - {\alpha \over {8\pi f_\pi}} \pi^0
  \epsilon^{\mu \nu \lambda \sigma} F_{\mu \nu} F_{\lambda \sigma},
 \label{point-anomaly}
\end{equation}
where $ f_\pi \simeq 93 {\rm MeV} $ is the pion decay constant
and $ F_{\mu\nu} $ is the field strength of photon,
describes
the behavior of $ \pi^0 \gamma \gamma $ vertex
in the limit of zero pion momentum and on-shell photons.
 However, naive use of Eq. (\ref{point-anomaly})
for the $ \pi^0 \gamma \gamma $ vertices in the diagram
of Fig. \ref{fig:diagrams}(b) leads to an ultra-violet divergent
result.
 This is a signal that
the interaction (\ref{point-anomaly}) is not applicable
to photons and pions far off mass-shell
and must be replaced there by some form factor.  
 In Ref. \cite{K-had} such a form factor was introduced
by an $ad~hoc$ adoption of the VMD picture.
 Correcting a sign error in the previous calculation \cite{K-had},
this contribution was found to be
\begin{eqnarray}
 a_\mu(b) &=&
  \displaystyle{
   - 0.044\ 36\ (2) \times
   \left( \frac{\alpha}{\pi} \right)^3
  } 
  \nonumber \\
 &=&
  \displaystyle{
   - 55.60\ (3) \times 10^{-11} 
  } .
 \label{KNO-pion-pole}
\end{eqnarray}

 In the previous analysis the quark loop diagram
in Fig. \ref{fig:diagrams}(c)
has been treated as {\it not} independent of the first two diagrams. 
 Rather it was used
as an alternative approximation
of the hadronic light-by-light scattering contribution
to $ a_\mu $.
 If this assertion is correct,
the result for quark loop \cite{K-had}
\begin{eqnarray}
 a_\mu(c)
  &=&
  \displaystyle{
   0.048\ (3) \times 
   \left( \frac{\alpha}{\pi} \right)^3
  } ,
  \nonumber \\
  &=&
  \displaystyle{
   62\ (3) \times 10^{-11}
  } ,
 \label{KNO-q-loop}
\end{eqnarray}
in which constituent quark masses are used,
should be nearly equal to
the sum of (\ref{KNO-loop}) and (\ref{KNO-pion-pole}).
However, even their signs do not agree with each other.
 
 Therefore there arises the second question:

($Q_2$) Are three diagrams shown in Fig. \ref{fig:diagrams}
    independent after all ? 

 We will examine these questions 
and explore the prescriptions for remedy 
in the next subsection.

\subsection{Improvements}
\label{subsec:improvement}

 First consider the question $Q_1$.
 It has been pointed out
that the naive VMD model of \cite{K-had} does not respect
the Ward identities required from electromagnetic gauge symmetry
\cite{Einhorn}.
 We found further that it is not compatible with chiral symmetry.
 To solve these problems, 
it is useful to introduce VMD in a way that preserves chiral symmetry.
 This can be achieved by appealing to
the hidden local symmetry (HLS) approach \cite{Bando}.
 This formulation maintains gauge invariance
and chiral symmetry explicitly
and reproduces all the low energy theorems assured by chiral symmetry,
such as the KSFR relation.
The question $Q_1$ may thus be raised within the HLS framework.  
 Keep in mind, however, that this approach is somewhat oversimplified.
 In particular it ignores higher resonances
beyond the usual vector mesons.
 We must analyze and reevaluate the error 
in our final result 
taking account of the model-dependence.


 We shall now turn to the second question $Q_2$.
 The previous work assumed that
the quark loop calculation and the pion calculation are
two distinct approximations to
the same hadronic light-by-light scattering effect on $ a_\mu $.
 They should, therefore, yield the similar results
and must not be added together.
As was noted in Ref. \cite{deRafael}, however,
in the extended Nambu-Jona-Lasinio (ENJL) model,
the quark loop diagram contribution
is independent of the other two
so that all three contributions should be added altogether.

 This point can be made clearer
by considering the $ 1/N_c $ expansion
together with the chiral expansion.
 Table \ref{tab:orders} lists the orders of each diagram
shown in Fig. \ref{fig:diagrams}.
 According to the QCD-diagrammatic consideration,
the pion loop needs at least two quark loops,
and the pion pole diagram starts from a diagram
in which at least one gluon propagates
between a quark and an antiquark forming the pion.
 Thus a single quark loop contribution is not included
in these two graphs.

 We may also examine this problem from the viewpoint of duality.  
 In a dispersion
relation for the light-by-light scattering amplitude,
the integral
over the quark loop diagram,
from threshold to very high energies, is equal 
to the integral over the absorptive part 
due to all the hadronic intermediate states.
 Extension of this relation
to local duality implies that the quark loop contribution
approximates
the hadronic contribution when certain averaging over a finite
energy region is taken.
 Thus one may wonder if the quark loop
diagram Fig. 2(c),
when embedded in the $g - 2$ graph,
represents the entire hadronic contribution. 

 We do not think so for the following reason.
 Consider a similar problem
in the photon vacuum polarization diagram, and look at
the cross section for low energy reaction
$e^+e^- \rightarrow hadrons$.
  Duality between the quark diagram and the resonances implies
  that
\begin{equation}
 \int \sigma(e^+e^- \rightarrow hadrons ; Q^2)dQ^2 \approx 
 \int \sigma(e^+e^- \rightarrow q\bar q ; Q^2)dQ^2 
\end{equation}
where the integration range is sufficiently large
so that the resonance is averaged out.
In this sense, it is well known that 
the absorptive part of one quark loop diagram
for the photon vacuum polarization is dual to
the cross section
from the $\pi\pi$ threshold to about 1 GeV.
 This region is dominated by the $\rho$ meson intermediate
state, and the quark loop represents this contribution. 
 Then there is a small continuum $\pi\pi$ state contribution
at lower energies. 
 Near threshold, they can be calculated
by chiral symmetry argument.
 Is this a part of the quark loop diagram
or is it a non-resonating continuum? 
 If it is a part of $\rho$, we suspect that a chiral 
 invariant $\pi\pi$ interaction
should be able to generate
the $\rho$ meson bound state. 
 As is well known \cite{Atkin}, however,
the force between two $\pi$ mesons
is not attractive enough to generate a bound $\rho$
state.
 This supports the view that the $\pi\pi$ intermediate state 
at low energies is independent of $\rho$ resonance
and hence independent
of one-quark-loop contribution.
 For the photon vacuum polarization graph,
pion loop graph and quark loop graph are independent and
must be added together.

 The above argument suggests that the pion loop
is independent of
the quark loop in light-by-light scattering, too. 
 The quark loop diagram corresponds to the sum of
continuum hadronic channels
as well as axial vector meson states.
 Of course, this is by no means a proof.
 Since it is impossible to prove this type of statements
without solving QCD,
we must keep this ambiguity in mind in our subsequent analysis.

 Let us summarize the above considerations and add a few corollaries:

 (1)
The HLS approach avoids the inconsistency that has been observed
in the naive VMD approach used in the previous analysis,
i.e., violation of chiral symmetry
and electromagnetic Ward identities
(This will be demonstrated in Sec. \ref{subsec:identity}).

 (2)
 Three diagrams shown in Fig. \ref{fig:diagrams} should be added.
 Especially the quark loop diagram
which represents the averaged hadronic continuum effect
in a certain energy region has been discussed as independent of
the other two.

 (3)
 Contributions involving more loops of hadrons
will be suppressed
by a factor $ m_\mu /(4\pi f_\pi ) $
compared to the contributions
of Fig. \ref{fig:diagrams}(a) and Fig. \ref{fig:diagrams}(b).
 On the other hand, the $ \eta $ pole contribution
from a diagrams similar to that of 
Fig. \ref{fig:diagrams}(b) may not be negligible.
 The magnitude of this contribution 
and the kaon loop contribution
from diagrams similar to Fig. \ref{fig:diagrams}(a)
deserves an explicit analysis.

 (4)
 As was mentioned already, naive use of Eq. (\ref{point-anomaly})
for the $ \pi^0 \gamma \gamma $ vertices 
of Fig. \ref{fig:diagrams}(b) leads to ultra-violet divergence,
indicating that (\ref{point-anomaly})
must be modified by a form factor far off mass-shell.
 Possible modification dictated
by the asymptotic behavior of QCD will be discussed in Sec. VI.
 Here we simply note
that the prescription adopted in Ref. \cite{K-had},
in which the VMD was introduced merely as a convenient UV cut-off, 
can be justified within the HLS approach \cite{Harada}.
 Note that the HLS Lagrangian 
 can be obtained as an effective theory of 
 the ENJL model
\cite{Ebert}.
Thus, a similar conclusion can also be reached 
in the ENJL model
 - the pion pole diagram contains two triangle loops
of constituent quarks
and $ \rho $ mesons is allowed to propagate
before the quarks couple to photons.
 Fig. \ref{fig:pi0-VMD} shows this contribution diagrammatically.
 However, its evaluation needs some care,
especially due to the requirement of anomalous Ward identities
\cite{Bijnens},
as is described in Sec. \ref{subsec:pole-2}.

 (5)
 In the quark loop diagram,
the vector meson will affect the coupling of the quark to the photon.
 Using the ENJL model as a guide we determine the quark coupling
to vector mesons.
 Its graphical expression is found
in Fig. 4 of Ref. \cite{deRafael}.

\section{Charged Pseudoscalar Loop}
\label{sec:loop}

\subsection{Hidden Local Symmetry Approach}
\label{subsec:HLS}

 For a complete description of HLS,
the reader is referred to Ref.\cite{Bando}.  
 Mainly for the purpose
of giving the Feynman rule relevant to our problem,
we shall briefly 
discuss the formalism. 
 The HLS incorporates
vector mesons, such as $ \rho $, 
as gauge particles of HLS,
$ [SU(2)_V]_{\rm local} $ in our case.
 The explicit form of the Lagrangian, assuming
chiral symmetry
$ \left[ SU(2)_L \times SU(2)_R \right]_{\rm global} $
and hidden local symmetry $ [SU(2)_V]_{\rm local} $ is
\begin{eqnarray}
 {\cal L} &=&
  \displaystyle{
   - \frac{1}{2 g_V^2} \tr\left( F_{V\,\mu\nu} F_V^{\mu\nu} \right)
   - \frac{1}{4}  F_{\mu\nu} F^{\mu\nu}
   + {\cal L}_A + a {\cal L}_V
  }
  \nonumber \\
  &&
  \displaystyle{
   + \frac{1}{2}
     f_\pi^2 B_0
       \left\{
        \tr( \xi_L {\cal M} \xi_R^\dagger )
                + \tr( \xi_R {\cal M}^\dagger \xi_L^\dagger )
       \right\}
  }
 \label{hidden}
\end{eqnarray}
where
\begin{eqnarray}
 {\cal L}_A &=& f_\pi^2
  \tr \left( (\hat{\alpha}_\parallel^\mu(x) )^2 \right),
  \nonumber \\
 {\cal L}_V &=& f_\pi^2
  \tr \left( (\hat{\alpha}_\perp^\mu(x) )^2 \right).
\end{eqnarray}
 In Eq.(\ref{hidden})
$ F_{V\,\mu\nu} $ and $ F_{\mu\nu} $ are
the field strengths of the vector meson
$ V_\mu = g_V \frac{\tau^a}{2}V^a_\mu  $
($ \tau^a , a = 1,2,3 , $ are Pauli matrices) and
the photon $ A_\mu $, respectively,
and $ g_V $ represents
the coupling constant associated with HLS.
$f_{\pi}$ is the pion decay constant ($\sim$ 93 MeV),
and the coefficient $a$ of $ {\cal L}_V $
is an arbitrary constant to be fixed by experiment.  
 $ \hat{\alpha}_{\{\parallel,\perp\}\,\mu} $ consists
of covariant derivatives of the basic objects
$ \xi_L(x) $ and $ \xi_R(x) $
in the HLS approach:
\begin{eqnarray}
 \hat{\alpha}_{\parallel\mu} &=&
 \displaystyle{
  \frac{D_\mu \xi_L \cdot \xi_L^\dagger
         + D_\mu \xi_R \cdot \xi_R^\dagger}{2i} ,
 }
  \nonumber \\
 \hat{\alpha}_{\perp\mu} &=&
 \displaystyle{
  \frac{D_\mu \xi_L \cdot \xi_L^\dagger
         - D_\mu \xi_R \cdot \xi_R^\dagger}{2i},
 }
\end{eqnarray}
where the covariant derivatives $ D_\mu \xi_{L,R}(x) $ are given by
\begin{equation}
 D_\mu \xi_{L,R}(x) =
  \del_\mu \xi_{L,R}(x) - i V_\mu(x) \xi_{L,R}(x)
   + i e \xi_{L,R}(x) \frac{\tau^3}{2} A_\mu(x).
\end{equation}
 $ \xi_L $ and $\xi_R$ contain the pion field $ \pi^a(x) $
as well as
the scalar triplet $ \sigma^a(x) $:
\begin{eqnarray}
 \xi_R(x) &=&
  \displaystyle{
   e^{i\sigma(x)/ f_\pi} e^{i\pi(x)/ f_\pi},
  }
 \nonumber \\
 \xi_L(x) &=&
  \displaystyle{
   e^{i\sigma(x)/ f_\pi} e^{-i\pi(x)/ f_\pi},
  }
\end{eqnarray}
where $ \pi(x) = \pi^a(x) \frac{\tau^a}{2} $
and $ \sigma(x) = \sigma^a(x) \frac{\tau^a}{2} $.
The latter, on breaking the symmetry,
will be absorbed into vector mesons $ V_\mu $ to give them masses.
 In the last term of Eq. (\ref{hidden})
$ B_0 $ is a dimension-one constant associated
with the quark condensate \cite{Gasser}
\begin{equation}
 B_0 = - \frac{1}{f_\pi^2}
         \left< 0 \right| \bar{u} u \left| 0 \right>.
\end{equation} 
It combines with
the current quark mass $ m_u $
in the mass matrix $ {\cal M} = {\rm diag}(m_u, m_d) $
(we neglect the isospin violation
due to the quark masses so that we set $ m_d = m_u $ henceforth)
to give the pion masses
\begin{equation}
 m_{\pi^\pm}^2 = m_{\pi^0}^2 = 2 B_0 m_u.
\end{equation}

 In the unitary gauge $ \sigma = 0 $ for HLS,
the relevant interaction terms for the present computation
can be found as follows:
\begin{eqnarray}
 {\cal L}_{\rm int} &=&
 - e g_\rho A^\mu \rho^0_\mu
 - i g_{\rho\pi\pi} \rho^0_\mu
   \pi^+ \deldel^\mu \pi^-
 - i g_{\gamma\pi\pi} A_\mu
   \pi^+ \deldel^\mu \pi^-
 \nonumber \\
 &&
 + (1-a) e^2 A^\mu A_\mu \pi^+ \pi^-
 + 2 e g_{\rho\pi\pi} A^\mu \rho^0_\mu \pi^+ \pi^-.
\label{int}
\end{eqnarray}
 In this expression
various masses and coupling constants
are related to each other by \cite{Bando}
\begin{eqnarray}
 M_\rho^2 &=& a g_V^2 f_\pi^2,
  \\
 g_\rho &=& a g_V f_\pi^2,
  \label{grho}
  \\
 g_{\rho \pi\pi} &=&
  \displaystyle{ \frac{1}{2} a g_V },
  \\
 g_{\gamma \pi\pi} &=&
  \displaystyle{
   \left( 1 - \frac{a}{2} \right) e
  },
  \label{ppipicoupling}
\end{eqnarray}
and $ A_\mu $ represents the photon field to the order $ e^2 $.  
 As is seen from Eq. (\ref{ppipicoupling})
the complete vector meson dominance
(namely $ g_{\gamma \pi \pi} = 0 $)
is realized when $ a = 2 $.
This is also close to the observed data.  
 Note that Eq. (\ref{int}) does not contain
the $ \rho^0 \rho^0 \pi^+ \pi^- $ term.  
 This is the crucial difference
between the chiral Lagrangian (\ref{hidden})
and the VMD model of Ref. \cite{K-had}.
 The absence of $ \pi^+ \pi^- \rho^0 \rho^0 $ coupling with
no derivatives will be a common feature
of chiral symmetric effective model,
as implied by other models, too
\cite{Ecker,Kaymakcalan} .

\subsection{Ward Identity}
\label{subsec:identity}

%
 Einhorn argued \cite{Einhorn}
that the calculation of Ref. \cite{K-had} does not satisfy
the Ward identities among the couplings
of $ \pi $ and $ \gamma $ required
from the electromagnetic symmetry.
 The purpose of this subsection is to demonstrate its recovery
in the present approach.  
 For simplicity let us consider 
a $ \pi \gamma $ scattering amplitude
and show explicitly
that the relevant Ward identity is satisfied
in the present approach.
 If we define the amputated Green functions $G^{\mu \nu}$
and $\Gamma^\mu$ in momentum space by
\begin{eqnarray}
 &&
 (2\pi)^4 \delta^4 ( q-k+p_1-p_2 )
 G^{\mu\nu}( q,k;p_1,p_2 )
 =
 \nonumber \\
 && \quad
 \int d^4 z e^{i q \cdot z} \int d^4 x e^{-i k \cdot x}
 \int d^4 y_1 e^{i p_1 \cdot y_1} \int d^4 y_2 e^{-i p_2 \cdot y_2}
 \left< 0 \left|
  T \left[
     j_{em}^\mu(z) A^\nu(x) \pi^+(y_1) \pi^-(y_2)
    \right]
 \right| 0 \right>_{{\rm amp.}},
 \nonumber \\
 &&
 (2\pi)^4 \delta^4 ( k-p_1+p_2 )
 i \Gamma^\mu( k;p_1,p_2 )
 \nonumber \\
 && \quad =
 \int d^4 x e^{-i k \cdot x}
 \int d^4 y_1 e^{i p_1 \cdot y_1} \int d^4 y_2 e^{-i p_2 \cdot y_2}
 \left< 0 \left|
  T \left[
     A^\mu(x) \pi^+(y_1) \pi^-(y_2)
    \right]
 \right| 0 \right>_{{\rm amp.}},
\end{eqnarray}
and denote the (full) pion propagator as $ i D(p) $,
the Ward identity can be written as
\begin{eqnarray}
 - q^\mu G_{\mu\nu} (q,k;p_1,p_2)
 &=&
 \displaystyle{
   \frac{ i D(p_1+q) }{ i D(p_1) }
   \Gamma_\nu (k;p_1+q,p_2)
   - \frac{ i D(p_2-q) }{ i D(p_2) }
     \Gamma_\nu (k;p_1,p_2-q)
\label{Ward-identity}
 }.
\end{eqnarray}
 In the naive VMD model of Ref. \cite{K-had} the photon propagator
\cite{gauge-choice}
\begin{equation}
 \frac{-i}{p^2},
 \label{photon}
\end{equation}
is replaced everywhere by
\begin{equation}
 \frac{-i}{p^2} - \frac{-i}{p^2 - M_\rho^2}
 = i \frac{M_\rho^2}{p^2(p^2-M_\rho^2)}.
 \label{photon-rho}
\end{equation}
 Thus, in this VMD model,
the Green function  $ \Gamma_\mu $ (or $ G_{\mu\nu} $) defined above
is simply obtained
by the multiplication of one (or two) $ \rho $ propagator(s)
to the corresponding quantity in scalar QED
\begin{eqnarray}
 &&
 G_{\mu\nu}(q,k,p_1,p_2) =
 \displaystyle{
  - e \frac{M_\rho^2}{q^2 - M_\rho^2}
      \frac{M_\rho^2}{k^2 - M_\rho^2}
   \left[
     2  g_{\mu\nu}
     - 
    \frac{1}{(k+p_2)^2-m_\pi^2}
     (2 p_2 + k)_\nu
     (2 p_1 + q)_\mu
   \right.
 }
  \nonumber \\
 && \quad \quad \quad \quad \quad \quad \quad \quad \quad
  \displaystyle{
   \left.
    -
    \frac{1}{(p_1-k)^2-m_\pi^2}
     (2 p_1 - k)_\nu
     (2 p_2 - q)_\mu
   \right]
 },
   \nonumber \\
 &&
 \displaystyle{
  \Gamma_\mu(k;p_1,p_2) =
  e \frac{M_\rho^2}{M_\rho^2 - k^2}
  (p_1+p_2)_\mu
 }.
  \label{eq:n-VMD-WI}
\end{eqnarray}
 Evidently the identity (\ref{Ward-identity}) cannot
hold due to the difference in the numbers of $ \rho $ propagators
between $ \Gamma_\mu $ and $ G_{\mu\nu} $.
 On the other hand, in the HLS approach, they are given,
to the order of our interest, by
\begin{eqnarray}
 &&
 G_{\mu\nu}(q,k;p_1,p_2) =
 \displaystyle{
  - e
  \left[
   2 \left\{
      g_{\mu\nu} - \frac{a}{2} H_{\mu\nu}(k) - \frac{a}{2} H_{\mu\nu}(q)
     \right\}
  \right.
 } 
 \nonumber \\
 && \quad \quad
 \displaystyle{
  - \frac{1}{(k+p_2)^2-m_\pi^2}
   \left\{
    g_{\nu\beta} - \frac{a}{2} H_{\nu\beta}(k)
   \right\} (2 p_2 + k)^\beta
   \left\{
    g_{\mu\alpha} - \frac{a}{2} H_{\mu\alpha}(q)
   \right\} (2 p_1 + q)^\alpha
 }
 \nonumber \\
 && \quad \quad
 \displaystyle{
 \left.
  - \frac{1}{(p_1-k)^2-m_\pi^2}
    \left\{
      g_{\nu\beta} - \frac{a}{2} H_{\nu\beta}(k)
    \right\} (2 p_1 - k)^\beta
    \left\{
     g_{\mu\alpha} - \frac{a}{2} H_{\mu\alpha}(q)
    \right\} (2 p_2 - q)^\alpha
 \right],
 }
 \nonumber \\
 &&
 \displaystyle{
  \Gamma_\mu(k;p_1,p_2) =
  e
   \left(
    g_{\mu\beta} - \frac{a}{2} H_{\mu\beta}(k)
   \right) (p_1+p_2)^\beta
 }.
 \label{eq:HLS-WI}
\end{eqnarray}
where $ H_{\mu\nu}(k) $ is defined by
\begin{equation}
 H_{\mu\nu}(k) = \frac{1}{k^2 - M_\rho^2}
                 ( g_{\mu\nu} k^2 - k_\mu k_\nu ).
\end{equation}
 It is an easy algebraic task
to confirm that the identity (\ref{Ward-identity}) holds now.
 In order to consider the recovery  of Ward identity in more detail,
we may add the $ k_\mu k_\nu / M_\rho^2 $-term
to each $\rho$-meson propagator
in Eq. (\ref{eq:n-VMD-WI})
by treating it as massive vector meson
\begin{equation}
  g_{\mu\nu} \frac{M_\rho^2}{M_\rho^2 - k^2}
   \rightarrow
   \frac{M_\rho^2}{M_\rho^2 - k^2}
   \left( g_{\mu\nu} - \frac{k_\mu k_\nu}{M_\rho^2} \right)
   = g_{\mu\nu} - H_{\mu\nu}(k) .
\end{equation}
 Then the expression (\ref{eq:n-VMD-WI}) becomes
\begin{eqnarray}
 &&
 G_{\mu\nu}(q,k;p_1,p_2) =
 \displaystyle{
  - e
  \left[
   2 \left\{
      g_{\mu\nu} - H_{\mu\nu}(k) - H_{\mu\nu}(q)
      + H_{\nu\beta}(k) H^\beta_{\ \mu}(q)
     \right\}
  \right.
 } 
 \nonumber \\
 && \quad \quad
 \displaystyle{
  - \frac{1}{(k+p_2)^2-m_\pi^2}
   \left\{
    g_{\nu\beta} - H_{\nu\beta}(k)
   \right\} (2 p_2 + k)^\beta
   \left\{
    g_{\mu\alpha} - H_{\mu\alpha}(q)
   \right\} (2 p_1 + q)^\alpha
 }
 \nonumber \\
 && \quad \quad
 \displaystyle{
 \left.
  - \frac{1}{(p_1-k)^2-m_\pi^2}
    \left\{
      g_{\nu\beta} - H_{\nu\beta}(k)
    \right\} (2 p_1 - k)^\beta
    \left\{
     g_{\mu\alpha} - H_{\mu\alpha}(q)
    \right\} (2 p_2 - q)^\alpha
 \right],
 }
 \nonumber \\
 &&
 \displaystyle{
  \Gamma_\mu(k;p_1,p_2) =
  e
   \left(
    g_{\mu\beta} - H_{\mu\beta}(k)
   \right) (p_1+p_2)^\beta
 }.
  \label{eq:n-VMD-WI-2}
\end{eqnarray}
 The comparison of Eqs. (\ref{eq:n-VMD-WI-2}) and (\ref{eq:HLS-WI})
with $ a = 2 $
shows that the absence of
the $ H_{\beta\nu}(k) H^\beta_{\ \mu}(q) $ term
in (\ref{eq:HLS-WI}) is responsible for the recovery of
the identity (\ref{Ward-identity}).
 This is a consequence of the nonexistence of the direct
$ \rho^0 \rho^0 \pi^+ \pi^- $-term, as has been stressed
in Sec. \ref{subsec:HLS}.
 This argument applies equally well
to the light-by-light scattering amplitude
caused by a charged pion loop.  

\subsection{Muon Anomaly}
\label{sec:muon-anomaly}

 We can now evaluate contributions of diagrams of Fig. 2 
to the muon anomaly $ a_\mu $.
 Let the vertex correction
from a diagram $ S $ be denoted as $ \Lambda^\nu_S(p,q) $
for the incoming photon momentum $ q $,
apart from the factor "$ i e $".
 Then the contribution to $ a_\mu $
from the diagram $ S $ is given by
\begin{equation}
 a_\mu(S)
  = \mathop{\lim}_{p.q, q^2 \rightarrow 0}
    \tr \left( P_\nu (p,q) \Lambda^\nu_S (p,q) \right),
\end{equation}
where $ P_\nu(p,q) $ is the magnetic moment projection operator
\begin{equation}
 P_\nu(p,q) =
 \displaystyle{
  \frac{1}{16}
   \left( \ps - \frac{\qs}{2} + 1 \right)
   \left( \gamma_\nu \qs - \qs \gamma_\nu - 3 p_\nu q.q \right)
   \left( \ps + \frac{\qs}{2} + 1 \right)
 },
\end{equation}
with the muon mass $ m_\mu $ set equal to 1.
 The diagrams constructed from the interactions in Eq. (\ref{int})
can be classified
in the similar manner as in Fig. 5 of Ref. \cite{K-had}.
 However, let us recall that
the replacement (\ref{eq:sub})
performed in the VMD model of \cite{K-had} 
works only if the $ \rho^0 \rho^0 \pi^+ \pi^- $ coupling term is present.
 In a theory with HLS
( restricted for simplicity
to the case of complete vector meson dominance
( $ a = 2 $ ) ),
however, 
there is no such term.  
 This means that, while 
the replacement
of the photon propagator (\ref{photon}) by (\ref{photon-rho}) 
is performed as before   
if the photon line is connected to the pion through
the $ \gamma \pi^+ \pi^- $ coupling,
the replacement must be carried out for either one of the lines but
{\it not for both},
if the photon lines comes from
a $ \gamma \gamma \pi^+ \pi^- $ coupling. 
 As a consequence, 
the contribution $ a_\mu ({\rm HLS}; A_2) $ from the diagram
in Fig. \ref{fig:A2}
(which is topologically the same
as the diagram $ A_2 $ in Fig. 5 of Ref. \cite{K-had}),
for instance, 
takes the form
\begin{eqnarray}
 \displaystyle{a_\mu ({\rm HLS}; A_2)} &=&
 \displaystyle{
  a_\mu ({\rm sQED}; A_2)
  - a_\mu ({\rm sQED}; (A_2,2))
  - a_\mu ({\rm sQED}; (A_2,3))
  - a_\mu ({\rm sQED}; (A_2,4))
 } \nonumber \\
 &&
 \displaystyle{
  + a_\mu ({\rm sQED}; (A_2,\{ 2, 3 \}))
  + a_\mu ({\rm sQED}; (A_2,\{ 3,4 \}))
 },
\end{eqnarray}
where $ a_\mu ({\rm sQED}; (A_2,\{ 2,3, \cdots \})) $
denotes
the quantity obtained by replacing the photon propagators
of the lines
$ 2, 3, \cdots $ with the propagators of mass $ M_\rho $.
 This differs from the calculation of Ref. \cite{K-had}
by the absence of the terms
\begin{equation}
  + a_\mu ({\rm sQED}; (A_2,\{ 2,4 \} ))
  - a_\mu ({\rm sQED}; (A_2,\{ 3,2,4 \})).
  \label{diff}
\end{equation}
 Since the $\gamma \pi^+ \pi^- $ vertex receives no modification, 
the contributions of the diagrams $ C_1 - C_4 $ in Ref. \cite{K-had}
remain unaltered.

\par
 The prescription for numerical evaluation of
Feynman integrals follows
that described in Ref.\cite{K-numerical}.
 As in scalar QED, the $ B_{ij}^\prime $ which appears
on the right-hand-side of Eq. (37) of Ref. \cite{K-numerical}
must be changed to $ B_{ij} $.
 The correctness of this change can be shown explicitly
in the same manner as in the Appendix B of Ref.\cite{K-had}.
 The renormalization \cite{K-ren} is required
for calculating individual diagram
since each diagram, not being gauge-invariant,
has logarithmic divergence 
residing in the hadronic light-by-light scattering subdiagram.  
 The evaluation of integrals is performed
with the help of the Monte Carlo integration routine 
VEGAS \cite{Lepage}.

 Let us now summarize our results.
 To begin with we checked the scalar QED result in (\ref{KNO-sQED})
by writing new FORTRAN programs from the scratch.
 The result
(obtained using VEGAS with 40 million sampling points
per iteration and 60 iterations) is 
\begin{eqnarray}
  a_\mu(a, {\rm sQED})
   &=&
   \displaystyle{
    -0.035~57~(18) \left( \frac{\alpha}{\pi} \right)^3 
   } 
    \nonumber \\
   &=&
   \displaystyle{
    - 44.58~(23) \times 10^{-11}
   } ,
  \label{newsQED}
\end{eqnarray}
confirming the previous result in (\ref{KNO-sQED}) but,
of course, with much higher precision.  
 The new evaluation of the $\rho$-meson contribution
in the HLS approach yields
(for 40 million sampling points per iteration and 60 iterations) 
\begin{eqnarray}
 a_\mu(a, {\rm HLS})
  &=&
  \displaystyle{
   - 0.003~55~(12) \left( \frac{\alpha}{\pi} \right)^3
  } 
   \nonumber \\
 &=&
  \displaystyle{
   - 4.45~(15) \times 10^{-11}
  } .
 \label{HLSresult}
\end{eqnarray}
This result is about 3.5 times smaller than the VMD model result
given in (\ref{KNO-loop}).

 To see whether this reduction is real,
we have evaluated the difference
$ a_\mu(a, {\rm HLS}) - a_\mu(a, {\rm nVMD}) $ directly.  
 The result is 
(for 40 million sampling points per iteration and 50 iterations)
\begin{eqnarray}
 a_\mu(a, {\rm HLS}) - a_\mu(a, {\rm nVMD})  
 &=&
  \displaystyle{
   0.009~76~(4) \left( \frac{\alpha}{\pi} \right)^3
  } 
  \nonumber \\
 &=&
  \displaystyle{
   12.23\ (5) \times 10^{-11}
  } .
 \label{difference}
\end{eqnarray}
 From (\ref{HLSresult}) and (\ref{difference}) we obtain
\begin{eqnarray}
  a_\mu(a, {\rm nVMD})
  &=&
   \displaystyle{
    -0.013~36~(14) \left( \frac{\alpha}{\pi} \right)^3 
   } 
    \nonumber \\
  &=&
   \displaystyle{
    -16.74~(18) \times 10^{-11}
   } ,
 \label{newVMD}
\end{eqnarray}
which is consistent with (\ref{KNO-loop}).  
 Thus the numerical works check out
and the difference (\ref{difference}) is real.  
 Of course,
the errors quoted above are those of numerical integration only
and do not include estimates of model dependence.  

%
\subsection{Discussion of Large Momentum Contribution}
\label{subsec:discussion}
%

 As we have seen in the previous subsection,
the $ \rho $ dominance structure has significant effects
on the hadronic light-by-light scattering contribution to $ a_\mu $. 
 In order to gain some insight
in the dependence of $a_\mu(a; {\rm HLS})$ on $M_\rho$ and $m_\pi$, 
let us introduce a function $a_\mu(a; m, M)$, 
where $a_\mu(a; {\rm HLS})$ = $a_\mu(a; m, M)$
for $m = m_\pi$ and $M = M_\rho$,
and examine it as a function of $m$ and $M$ numerically.  
 Table \ref{tab:pi-mass} contains
the result for $ m_\pi $ ( $ M_\rho $ ) dependence
obtained by 15 ( 30 ) iterations of Monte Carlo integration
with one million sample points.
 From that table
the following approximate asymptotic behaviors
can be inferred:
\begin{eqnarray}
 &&
  \displaystyle{
   a_\mu(a; x m_\pi, M) \sim -4.75 \times 10^{-2} \times x^{-2}
    \left ( {\alpha \over \pi} \right )^3 
    \ \ \ \ {\rm for}
    \ \ \ \ x \geq 3 ,\  M = \infty
  } ,
 \label{asymp1} \\
 &&
  \displaystyle{
   a_\mu(a; x m_\pi, M) \sim +2.81 \times 10^{-2} \times x^{-2}
    \left ( {\alpha \over \pi} \right )^3 
    \ \ \ \ {\rm for}
    \ \ \ \ x \geq 3 ,\ M = M_\rho
  } ,
 \label{asymp2} \\
 &&
  \displaystyle{
   a_\mu(a; m_\pi, M) \sim a_\mu (a; m_\pi , \infty) +
    0.23 \left ( {m_\mu \over M } \right )
    \left ( {\alpha \over \pi} \right )^3 
    \ \ \ \ {\rm for}
    \ \ \ \ M > M_\rho
  } ,
 \label{asymp3}
\end{eqnarray}
 where $a_\mu(a; m_\pi , \infty ) = a_\mu(a, {\rm sQED}) $.  

 The results (\ref{asymp1}) and (\ref{asymp2}) show
that $a_\mu(a; m, M)$ depends on the loop mass $m$ 
asymptotically as $m^{-2}$ for a wide range of $`` \rho$ mass" $M$.  
 To appreciate the significance of these results,
note that, for $m >> m_\pi$, $m$
in $a_\mu(a; m_\pi, M) $ $-$ $a_\mu(a; m, M)$ may be
regarded as a cut-off mass
of the pion-loop momentum \`{a} la Pauli-Villars:  
 The contribution of pion-loop momenta above $m$ is
suppressed in
$a_\mu(a; m_\pi, M)$ $-$ $a_\mu(a; m , M)$.
 Thus the above dependences on $ m_\pi $ result from the fact
that the contribution of pion loop momenta larger than $m$
drops off as $m^{-2}$ as $m$ increases. 
 For instance,
the contribution of pion-loop momentum higher than 800 MeV
occupies only 7 percent
of the total contribution (\ref{newsQED}) or (\ref{HLSresult}).

 From these results, we speculate that
the pion-loop light-by-light scattering amplitude
(even with off shell photons)
is governed by the region
of small loop momenta carried by light hadrons.

 The result (\ref{asymp3}) is inferred from the near constancy of  
$M(a_\mu(a; m_\pi , M) - a_\mu(a; m_\pi , \infty ))$
for $M_\rho \leq M \leq 10 M_\rho$.  
 This function decreases very slowly for larger $M$. 
 Such an $M^{-1}$ (instead of $M^{-2}$) behavior
seems to cast some doubt on the effectiveness
of chiral perturbation theory
since it implies that there is an appreciable contribution to
$a_\mu(a, {\rm HLS})$
from the region of photon momenta larger than $M_\rho$.  
 To analyze this problem,
let us recall
that the customary argument in favor of the $M^{-2}$ behavior  
is inferred from the fact that $-M_\rho^{-2}$ is the dominant term in 
the $\rho$-meson propagator:
\begin{equation}
 \frac{1}{p^2 - M_\rho^2}
 = -\frac{1}{M_\rho^2} + \frac{p^2}{(p^2-M_\rho^2)M_\rho^2}
 \label{expand} 
\end{equation}
for $|p^2| \ll M_\rho^2$. 
 As is readily seen by power counting,
however, naive evaluation of the contribution of each term on the
right-hand side of Eq. (\ref{expand}) to $a_\mu(a, {\rm HLS})$
leads to UV cut-off dependent results.
 In other words,
the $M^{-2}$ term has a divergent coefficient
and naive power counting argument fails. 
 This is due to the fact
that the $M_\rho \rightarrow \infty$ limit
and sub-integrations in the Feynman diagram
do not commute. 

 It is important to note, however, that the fact that 
photons of momenta larger than $M_\rho$ contribute significantly
to $a_\mu(a, {\rm HLS})$ does not necessarily
mean the failure of chiral perturbation theory.  
 The requirement on the photon momentum,
which is a vector sum of pion momenta, can be considerably less strict  
insofar as the contribution to $a_\mu(a, {\rm HLS})$
comes mostly from small pion loop momenta.  
 In fact this is what can be inferred
from (\ref{asymp1}) and (\ref{asymp2}).
 For these reasons the $M^{-1}$ behavior of (\ref{asymp3})
is not inconsistent with the chiral perturbation theory.

 For $M = M_\rho$, (\ref{asymp3}) can be written as 
\begin{equation}
 a_\mu(a, {\rm HLS})
 \simeq \left ( -0.035\ 57\ (18) 
 + 0.23 {m_\mu \over M_\rho} \right)
 \left ( {\alpha \over \pi} \right)^3 ,
 \label{approx}
\end{equation}
where the first term is from (\ref{newsQED}).  
 $a_\mu(a, {\rm HLS})$ deviates from (\ref{approx})
for $M_\rho$ smaller than the physical value.  
 It is seen from (\ref{approx})
that the leading term happens to be nearly cancelled
by the non-leading term for physical $\rho$ mass.  
 The smallness of the value (\ref{HLSresult})
results from a (somewhat accidental) cancelation
of $a_\mu(a, {\rm sQED})$
and the ${\cal O} ( m_\mu / M_\rho )$ term
for the physical value of the $\rho$ meson mass.  

 Before writing down our best estimate of
the charged pion loop contribution
to the muon $g - 2$, 
it must be recalled that the result (\ref{HLSresult}) is based on
the specific hadron model.
 In principle, any models which preserve chiral symmetry
and the relevant Ward identities
are the candidates for this computation.
 Any of these models is expected to lead
to more or less
the same hadronic light-by-light scattering amplitude
provided that the chiral symmetry is intact.
 Eq. (\ref{approx}) indicates, however,
that the large photon momenta give rise to
significant contribution to $ a_\mu $.
 Thus the hadronic structure in photon beyond the $ \rho $ mass
may be non-negligible.
 The model dependence may enter here.

 In this computation, we have used the HLS approach.
 Even within this framework we assume
further a complete $ \rho $ dominance.
 Also we could
have chosen the version of HLS with higher resonances,
such as $ A_1 $.
All these would increase the uncertainty of our result.

 The previous work based
on the ENJL model \cite{deRafael}
asserts
that the QED result of pion loop should be included
from the standpoint of systematic chiral expansion.
 As was mentioned previously,
the ENJL model Lagrangian
is always written in the form consistent with 
HLS.
 Then the pion loop contribution in that model reduces to
the one obtained here
when we approximate more complicated form factor of
$ \pi^+ \pi^- \gamma \gamma $ and $ \pi^+ \pi^- \gamma $ \cite{foot}.

 In spite of these model dependences,
we expect the total error to be within 20 \% of the difference
$ a_\mu(a, {\rm HLS}) - a_\mu(a, {\rm sQED}) $.
This is because integrations over the photon and muon momenta
are convergent in these diagrams
and hence the contribution
of large photon momenta does not distort
our picture of low energy pion loop too severely.

 The kaon loop contribution is found to be about 4 \%
of the pion contribution.
 Taking these error estimates
of the pion-loop and kaon-loop contributions into consideration,
we present 
\begin{eqnarray}
  a_\mu(a)
   &=&
    \displaystyle{
      - 0.003~6~(64) \left( \frac{\alpha}{\pi} \right)^3
    } 
     \nonumber \\
   &=&
    \displaystyle{
     -4.5\ (8.1) \times 10^{-11} 
    } ,
 \label{bestestimate}
\end{eqnarray}
as our best estimate, including model dependence,
for the contribution to $ a_\mu $ of the charged pion loop part of 
the hadronic light-by-light scattering amplitude.
 Eq. (\ref{bestestimate}) replaces
(\ref{KNO-loop}) obtained
in Ref. \cite{K-had} based on a naive vector meson dominance model
which is inconsistent with the chiral symmetry.

\section{Neutral Pseudoscalar Pole}
\label{sec:pole}

 The first subsection here describes the detail of
our prescription adopted in Ref. \cite{HKS}.
 The second subsection
describes extension of our method to include 
the pole-type axial contribution,  
and discusses the total contribution of this type.

\subsection{Incorporation of Triangle Quark Loop}
\label{sebsec:pole-1}

 For the purpose of later comparison,
we record again the result (\ref{KNO-pion-pole})
for the neutral pion pole
contribution $ a_\mu(b) $ obtained in the HLS approach
(for 5 million sampling points per iteration and 20 iterations):
\begin{eqnarray}
 a_\mu(b,{\rm HLS}) &=&
  \displaystyle{
   - 0.044\ 36\ (2)
   \left( \frac{\alpha}{\pi} \right)^3
  } 
   \nonumber \\
  &=&
   \displaystyle{
    - 55.60~(3) \times 10^{-11}
   } .
 \label{local-pi0}
\end{eqnarray}
 Here we used the newly written FORTRAN programs
for evaluating this result.
 Note that a sign error in some part
of the integrand in \cite{K-had} is corrected in (\ref{local-pi0}).

 It is far from certain that the off-shell behavior, in particular,
with respect to the pion momentum, is well-approximated
by the use of the effective interaction (\ref{point-anomaly})
modified by the HLS method.
 The examination of different off-shell extrapolation scheme
will give some insight
in the dependence of muon anomaly $ a_\mu(b) $
on the off-shell behavior.
 Here we choose the diagram shown in
Fig. \ref{fig:pi0-VMD},
which is again suggested by the ENJL model,
as a model for such an extrapolation scheme,
and evaluate it explicitly.

 Let us begin by noting that, in Fig. 3, the one-quark-loop subdiagram 
corresponding to $ \pi^0(q) \rightarrow \gamma(p_1) \gamma(p_2) $
can be written
as \cite{Jackiw}
\begin{equation}
  A_{\mu\nu} \left( p_1, p_2 \right)
   = \epsilon_{\mu \nu  \alpha \beta}
      p_1^\alpha p_2^\beta  { {\alpha} \over {\pi f_\pi }}
      \int[dz]
       {{2 m_q^2 } \over
        {m_q^2 - z_2 z_3 p_1^2 - z_3 z_1 p_2^2 - z_1 z_2 q^2 }} ,
    \label{triangle}
\end{equation}
where $ [dz] = dz_1 dz_2 dz_3 \delta(1-(z_1+z_2+z_3)) $.
 This amplitude is reduced to the one obtained from (\ref{point-anomaly})
in the limit $ p_1^2 = p_2^2 = q^2 = 0 $,
showing that it is normalized correctly.
 Note that the insertion of (\ref{triangle}) into the Feynman diagram
of $ a_\mu $ yields a convergent result
without recourse to the VMD model.
 If we use the notation $ a_\mu(b; m_\pi, M_\rho, m_q) $
in analogy with $ a_\mu(a; m_\pi, M_\rho) $ in the case of $a_\mu (a)$,
such a contribution corresponds to $ a_\mu(b; m_\pi,\infty, m_q) $.
 The numerical evaluation can be carried out
in a straightforward manner
if we re-express (\ref{triangle}) in the form of
momentum integral representation and use the general formalism
\cite{K-numerical} for the evaluation of Feynman integral.
 The result is found to be
(for 5 million sampling points per iteration and 20 iterations)
\begin{eqnarray}
 a_\mu(b; m_\pi,\infty, m_q) &=&
  \displaystyle{
   -0.069\ 34\ (5)
   \left( \frac{\alpha}{\pi} \right)^3
  } 
   \nonumber \\
  &=&
   \displaystyle{
    -86.90\ (7) \times 10^{-11}    \label{twoanomalyloop}
   } ,
\end{eqnarray}
where we have chosen the quark mass to be 300 MeV.
  The operator product expansion analysis
on the short distance behavior of
the amplitude $ A_{\alpha\beta} $ supports the use of constituent
quark mass as $ m_q $ in Eq. (\ref{triangle}) \cite{Manohar}.

 The diagram in Fig. \ref{fig:diagrams}(b),
with the VMD assumption added, can be
calculated in a similar manner.
 As has been noted in Sec. \ref{subsec:improvement},
the coupling of quark to vector meson is obtained
with the help of the ENJL model.
 As a result we obtain
(for 5 million sampling points per iteration and 20 iterations)
\begin{eqnarray}
  a_\mu(b) &=&
  \displaystyle{
    - 0.026\ 94(5) 
    \left( \frac{\alpha}{\pi} \right)^3
   } 
    \nonumber \\
   &=&
    \displaystyle{
     - 33.76~(7) \times 10^{-11 }
    } .
 \label{pi0-VMD-triangle}
\end{eqnarray}

 Here  again, in order 
to examine what range of momentum governs $ a_\mu(b) $,
we perform the same analysis on $ a_\mu(b; m_\pi, M_\rho, m_q) $,
where $m_q$ is constituent quark mass of the triangular loop, 
as has been done for $ a_\mu(a; m_\pi, M_\rho) $. 
 Table \ref{tab:pi0-mass} lists
the results for the quoted quantity
for various values of $ m_\pi $ ( or $ M_\rho $ )
obtained by 20 ( or 10 ) iterations of integration
with 5 million sample points.
 For quark mass larger than 300 MeV,
$ a_\mu(b; m_\pi, M_\rho, m_q) $
approaches $ a_\mu(b;{\rm HLS}) $ as $ m_q^{-2} $,
as is readily seen 
from the analytic expression for the triangle graph.
 For small $ m_q $, it approaches zero.
For instance, for $m_q$ = 5 MeV, we have 
$ a_\mu(b; m_\pi, M_\rho, m_q ) 
= -0.666 \times 10^{-5}(\alpha / \pi )^3 $.
 The results in Table \ref{tab:pi0-mass}
are summarized by the following asymptotic form:
\begin{eqnarray}
 &&
  \displaystyle{
   a_\mu(b; x m_\pi, M_\rho, m_q)
    = -9.57 \times 10^{-2} \times x^{-2}
    \left( \frac{\alpha}{\pi} \right)^3
    \ \ \ \ {\rm for} \ \ \ \ x \geq 3
  },
   \label{mpi-for-pi0} \\
 &&
  \displaystyle{
   a_\mu(b; m_\pi, M, m_q)
    = a_\mu(b; m_\pi,\infty, m_q)
      + 0.31 \left( \frac{m_\mu}{M} \right)
             \left( \frac{\alpha}{\pi} \right)^3
    \ \ \ \ {\rm for} \ \ \ \ M \geq 3 M_\rho
  }.
   \label{mrho-for-pi0}
\end{eqnarray}
 These results show that the same consideration
as in Sec. \ref{subsec:discussion}
also applies here.

 Note that the $ \eta $ pole contribution, when 
the mixing among $\pi$, $\eta$ and $\eta^\prime$
is taken into account,
amounts to 25 \% of (\ref{pi0-VMD-triangle}):
\begin{eqnarray}
 a_\mu({\eta}\ {\rm pole})
 &=&
  \displaystyle{
   - 0.005\ 29\ (2) \times
   \left( \frac{\alpha}{\pi} \right)^3
  }
   \nonumber \\
 &=&
  - 7.305\ (3) \times 10^{-11}  ,
\label{eta-pole-1}
\end{eqnarray}
which is obtained using 5 million sampling points per iteration
and 20 iterations.
The $ \rho $ mass dependence of the $ \eta $ contribution
is listed in Table \ref{tab:eta} 
(for 5 million sampling points per iteration
and 10 iterations). 
 From this Table we obtain an approximate asymptotic formula
\begin{equation}
 a_\mu(b;m_\eta,M,m_q) = a_\mu(b;m_\eta,\infty,m_q)
  + 0.11 \left( \frac{m_\mu}{M} \right)
    \times \left( \frac{\alpha}{\pi} \right)^3
  \ \ \ \ {\rm for}\ \ \ \
  M \ge 3 M_\rho, 
\end{equation}
where
\begin{eqnarray}
 a_\mu(b;m_\eta,\infty,m_q) &=&
  - 0.020\ 08\ (1) \times
  \left( \frac{\alpha}{\pi} \right)^3 ,
   \nonumber \\
 &=&
   - 25.17\ (2) \times 10^{-11} .
\end{eqnarray}
This was obtained
for 5 million sampling points per iteration and 20 iterations.

Adding (\ref{pi0-VMD-triangle}) and (\ref{eta-pole-1}) 
and again estimating the model dependence
to be within about 20 \% of the $ M_\rho $-dependent term, 
we obtain
\begin{eqnarray}
 a_\mu(b) &=&
 \displaystyle{
   -0.032\ 2\ (66)
   \left( \frac{\alpha}{\pi} \right)^3
  } 
   \nonumber \\
  &=&
   \displaystyle{
    -40.4\ (8.3) \times 10^{-11}
   } .
 \label{neutral-pole}
\end{eqnarray}
 This is the result
for the pseudoscalar pole contribution given in Ref.
\cite{HKS}.

 Further discussion of the off-mass-shell behavior
of the $\pi^0 \gamma^* \gamma^*$ vertex is given in Sec. VI.

\subsection{Further Examination of the Pole Contribution}
\label{subsec:pole-2}

 After our summarizing paper, Ref. \cite{HKS},
was submitted for publication, 
we learned that the hadronic light-by-light scattering contribution
to the muon $ g-2 $ has also been studied
by another group \cite{Bijnens}
in the large $ N_C $ limit within the framework of the ENJL model.
 Their initial result for the contribution corresponding
to Fig. 2(b) disagreed strongly with our result.  
 Since then, however,
it was found that this was due
to a simple numerical oversight \cite{Bijnens2}.
 Correction of this error brings
their result closer to our $a_\mu (b)$.
 Nevertheless, their report \cite{Bijnens}
stimulated our interest
to study the axialvector pole contribution in some detail.

 In order to facilitate comparison
with the results of \cite{Bijnens}, we follow their method closely.
 In particular 
we use the notation which enables us to keep track of
the normalization factor directly.
 For the operator $ j_5 \equiv \bar{q} T^3 i \gamma_5 q $,
$ j^Q_\mu \equiv \bar{q} Q \gamma_\mu q $,
where the isospin generator $ T^3 $ is normalized
as $ 2 {\rm tr}(T^3 T^3) = 1 $,
the PVV (pseudoscalar-vector-vector)
three-point function is defined as
\begin{eqnarray}
 \Pi^{\rm PVV}_{\mu\nu}(p_1,p_2)
  &\equiv&
  \displaystyle{
   i^2 \int d^4 x_1 e^{ip_1\cdot x_1}
       \int d^4 x_2 e^{ip_2\cdot x_2}
      \left< 0 \right| T j_5(0) j_\mu^Q(x_1) j_\nu^Q(x_2)
      \left| 0 \right>
  }.
\end{eqnarray}
The part of  
 $\Pi^{\rm PVV}_{\mu\nu}(p_1,p_2)$ 
corresponding to the 1-loop contribution,
$ \bar{\Pi}^{\rm PVV}_{\mu\nu}(p_1,p_2) $,
is given by the well-known triangle loop graph
\begin{eqnarray}
 &&
  \displaystyle{
   \bar{\Pi}^{\rm PVV}_{\mu\nu}(p_1,p_2) =
   - \frac{2}{m_q} \frac{1}{16\pi^2}
     \epsilon_{\mu\nu\alpha\beta} p_1^\alpha p_2^\beta
      F(p_1^2,p_2^2,q^2)
  },
   \nonumber \\
 &&
  \displaystyle{
    F(p_1^2, p_2^2, q^2) =
    1 + I_3(p_1^2, p_2^2, q^2) - I_3(0,0,0)
   },
    \nonumber \\
 &&
  \displaystyle{
   I_3(p_1^2,p_2^2,q^2) =
   2 m_q^2 \int d z_1 d z _2 d z_3 \delta(z_1+z_2+z_3 - 1)
    \frac{\Gamma(1,\tilde{M}^2(z_\alpha)/\Lambda_\chi^2)}
         {\tilde{M}^2(z_\alpha)}
  }, \nonumber \\
 &&
  \displaystyle{
    \tilde{M}^2(z_\alpha) =
    m_q^2 - p_1^2 z_2 z_3 - p_2^2 z_3 z_1 - q^2 z_1 z_2
   } ,
 \label{PVV-form}
\end{eqnarray}
where $ \Lambda_\chi $ is the momentum-cutoff which renders
the quark loop contribution finite,
$ m_q $ is the constituent quark mass,
and $ \Gamma(n,x) $ is the incomplete gamma function
\begin{equation}
 \Gamma(n,x) =
   \int_x^\infty dt e^{-t} t^{n-1} .
\end{equation}

 The leading $ 1/N_C $ term of 
$ \Pi^{\rm PVV}_{\mu\nu}(p_1,p_2) $ can be written as \cite{Prades}
\begin{eqnarray}
 \Pi^{\rm PVV}_{\mu\nu}(p_1,p_2)
  &=&
   \displaystyle{
    - \frac{1}{16\pi^2}
    \epsilon_{\mu\nu\alpha\beta} p_1^\alpha p_2^\beta
    \frac{4m_q}{g_S f_\pi^2(-q^2) \left( m_\pi^2(-q^2) - q^2 \right)}
   }
    \nonumber \\
  && \times
   \displaystyle{
    \left[
     1 - g_A(-q^2) \left\{
                     1 - F(p_1^2,p_2^2,q^2)  L(p_1^2,p_2^2)
                   \right\}
    \right]
   } ,
 \label{eq:PVV}
\end{eqnarray}
where
\begin{equation}
 L(p_1^2,p_2^2) = 
     \frac{M_V^2(-p_1^2) M_V^2(-p_2^2)}
          {\{ M_V^2(-p_1^2) - p_1^2\} \{ M_V^2(-p_2^2) - p_2^2\}}.
\end{equation}
and
$ g_S = 8\pi^2 G_S/{N_C \Lambda_\chi^2} $ is 
the scalar coupling constant in the ENJL Lagrangian
\begin{eqnarray}
 {\cal L}_{{\rm ENJL}}^{{\rm (int.)}}
  &=&
  \displaystyle{
   \frac{8\pi^2 G_S}{N_C \Lambda_\chi^2}
   \sum_{i,j} (\bar{q}^i_R q_{L\,j}) (\bar{q}^j_L q_{R\,i})
  }
   \nonumber \\
  &&
  \displaystyle{   
   - \frac{8\pi^2 G_V}{N_C \Lambda_\chi^2}
     \sum_{i,j}
      \left\{
       (\bar{q}^i_L \gamma^\mu q_{L\,j})
        (\bar{q}^j_L \gamma_\mu q_{L\,i})
       + (\bar{q}^i_R \gamma^\mu q_{R\,j})
        (\bar{q}^j_R \gamma_\mu q_{R\,i})
      \right\}
  }.
\end{eqnarray}
Here $ i $ and $ j $ represent the flavor indices,
$ N_C $ is the number of colors and the parentheses assume
the implicit sum over colors.
 The definition of various functions appearing in Eq.(\ref{eq:PVV})
can be found in Ref. \cite{Prades}.

 Next we turn our attention to
the AVV (axialvector-vector-vector) three-point function
defined by
\begin{equation}
 \Pi^{\rm AVV}_{\alpha\mu\nu}(p_1,p_2)
  = i^2 \int d^4 x_1 e^{ip_1\cdot x_1}
        \int d^4 x_2 e^{ip_2\cdot x_2}
        \left< 0 \right| T j_{5\alpha}(0) j_\mu^Q(x_1) j_\nu^Q(x_2)
        \left| 0 \right> .
\end{equation}
where $ j_{5\mu} \equiv \bar{q} \gamma_\mu \gamma_5 T^3 q $.  
 A direct evaluation gives
\begin{eqnarray}
 \Pi^{\rm AVV}_{\alpha\mu\nu}(p_1,p_2) &=&
 \displaystyle{ L(p_1^2, p_2^2)
   \left[
     g_A(-q^2) \frac{M_A^2(-q^2)}{M_A^2(-q^2) - q^2}
     \bar{\Pi}^{\rm AVV}_{\alpha\mu\nu}(p_1,p_2)
   \right.
  }
   \nonumber \\
 && \quad \quad \quad \quad
  \displaystyle{
     - 2 m_q
        g_A(-q^2) \frac{1}{M_A^2(-q^2) - q^2}
          i q_\alpha \bar{\Pi}^{{\rm PVV}}_{\mu\nu}(p_1,p_2)
  }
   \nonumber \\
  && \quad \quad \quad \quad
   \displaystyle{
    \left.
     + 2m_q g_A(-q^2) \frac{1}{m_\pi^2(-q^2) - q^2}
       \ i q_\alpha \bar{\Pi}^{{\rm PVV}}_{\mu\nu}(p_1,p_2)
    \right]
   }
    \nonumber \\
  &&
   \displaystyle{
    + 2m_q
         g_A(-q^2) \frac{1}{M_A^2(-q^2) - q^2}
         i q_\alpha
         \left.
            \bar{\Pi}^{{\rm PVV}}_{\mu\nu}(p_1,p_2)
         \right|_{p_1^2=0=p_2^2=q^2}
   }
    \nonumber \\
  &&
   \displaystyle{
    + 2m_q ( 1 - g_A(-q^2))
         \frac{1}{m_\pi^2(-q^2) - q^2}
         \ i q_\alpha
         \left.
           \bar{\Pi}^{{\rm PVV}}_{\mu\nu}(p_1,p_2)
         \right|_{p_1^2=0=p_2^2=q^2}
   },
 \label{eq:AVV}
\end{eqnarray}
where $ \bar{\Pi}^{{\rm PVV}}_{\mu\nu}(p_1,p_2)$ 
is the 1-loop contribution of 
$ {\Pi}^{{\rm PVV}}_{\mu\nu}(p_1,p_2)$.  
$ \bar{\Pi}^{\rm AVV}_{\alpha\mu\nu}(p_1,p_2) $ is a linearly
divergent integral
\begin{equation}
 \bar{\Pi}^{\rm AVV}_{\alpha\mu\nu}(p_1,p_2) =
 \frac{1}{2} i^2 2
  \int \frac{d^4 r}{(2\pi)^4} (-1)
  {{\rm tr}}
  \left[
   \gamma_\alpha \gamma_5 \frac{i}{\rs+\ps_1-m_q}
   \gamma_\mu \frac{i}{\rs-m_q}\gamma_\nu \frac{i}{\rs-\ps_2-m_q}
  \right],
\label{AVV}
\end{equation}
where Pauli-Villars regularization is understood.
 The last two terms of (\ref{AVV})
come from the presence of anomaly contribution
when $ -i q^\lambda \bar{\Pi}^{\rm AVV}_{\lambda\mu\nu}(p_1,p_2) $
is rewritten in terms of $ \bar{\Pi}^{\rm PVV}_{\mu\nu}(p_1,p_2) $:
\begin{eqnarray}
 \displaystyle{
  -i q^\lambda \bar{\Pi}^{\rm AVV}_{\lambda\mu\nu}(p_1,p_2)
  L(p_1^2,p_2^2)
 }
 &=&
 \displaystyle{
  2m_q
     \left\{
      \bar{\Pi}^{\rm PVV}_{\mu\nu}(p_1,p_2) L(p_1,p_2)
        - \left.
           \bar{\Pi}^{\rm PVV}_{\mu\nu}(p_1,p_2)
          \right|_{p_1^2 = p_2^2 = q^2 = 0}
     \right\}
 }, 
  \nonumber \\
\end{eqnarray}
a relation which was also used to derive Eq. (\ref{eq:PVV}).

 Now the contributions of pseudoscalar
and axial vector intermediate states
to the four-photon vertex graph
can be written, for instance, as
\begin{eqnarray}
 &&
  \displaystyle{
    (ie)^4
    \left[
     (2 i g_S) \Pi^{\rm PVV}_{\mu\nu}(p_1,p_2)
       \bar{\Pi}^{\rm PVV}_{\rho\sigma}(p_3,p_4) L(p_3^2, p_4^2)
    \right.
   } \nonumber \\
 && \quad \quad \quad
  \displaystyle{
    \left.
     +
     \left(
       - 2i \frac{8\pi^2 G_V}{N_C \Lambda_\chi^2}
     \right)
     \Pi^{\rm AVV}_{\alpha\mu\nu}(p_1,p_2)
     \bar{\Pi}^{\rm AVV\,\alpha}_{\ \ \ \ \ \ \rho\sigma}(p_3,p_4)
     L(p_3^2,p_4^2)
   \right]
  }. \nonumber \\
   \label{eq:pole-sum}
\end{eqnarray}

 The pseudoscalar pole contribution can be extracted
from (\ref{eq:pole-sum}):
\begin{eqnarray}
 i \hat{A}_{\mu\nu}(p_1,p_2)  \frac{i}{q^2 - m_\pi^2(-q^2)}
  i \hat{A}_{\rho\sigma}(p_3,p_4),
   \label{eq:ps-pole-contr}
\end{eqnarray}
where
\begin{eqnarray}
 \hat{A}_{\mu\nu}(p_1,p_2) &\equiv&
 \displaystyle{
   - \frac{8\pi\alpha m_q}{2f_\pi(-q^2)}
     \bar{\Pi}^{PVV}_{\mu\nu}(p_1,p_2)
 }
  \nonumber \\
 &=&
 \displaystyle{
  \frac{\alpha}{\pi f_\pi(-q^2)}
  \left[ 1 - g_A(-q^2)
             \left\{
               1 - F(p_1^2,p_2^2,q^2) L(p_1^2,p_2^2)
             \right\}
  \right]
 } .
  \label{eq:pseudo}
\end{eqnarray}

 For the following analysis,
momentum dependences of various functions will be ignored:
$ f_\pi^2(-q^2) \simeq f_\pi^2 $,
$ M_V^2(-q^2) \simeq M_\rho^2 $,
$ g_A(-q^2) \simeq g_A $,
$ q^2 - m_\pi^2(-q^2) \simeq A^2 (q^2 - m_\pi^2) $ with
$ A $ accounting for the wave function renormalization constant of pion
\begin{equation}
 A^2 = 1 - \left.
           \frac{\partial m_\pi^2(-p^2)}{\partial p^2}
           \right|_{p^2 = m_\pi^2},
\end{equation}
which is close ( and therefore set equal ) to unity,
and $ \Lambda_\chi $ is taken as $ \infty $.
 In this approximation Eq. (\ref{eq:ps-pole-contr})
reduces to
\begin{equation}
 iA_{\mu\nu}(p_1,p_2)  \frac{i}{q^2 - m_\pi^2} 
 iA_{\rho\sigma}(p_3,p_4),
\end{equation}
where
the amplitude shown in (\ref{triangle}),
multiplied by a function $ L(p_1^2, p_2^2) $
associated with vector meson dominance, 
is modified to
\begin{eqnarray}
 &&
  \displaystyle{
    A_{\mu\nu}(p_1,p_2) =
    \frac{\alpha}{\pi f_\pi} \epsilon_{\mu\nu\beta\rho}
    p_1^\beta p_2^\rho F_{{\rm PVV}}(p_1^2, p_2^2, q^2)
   } ,
 \label{eq:enjl-modify}
 \\
 &&
  \displaystyle{
   F_{{\rm PVV}}(p_1^2,p_2^2,q^2)  =
    \frac{1}{A}
      \left\{
         1 - g_A
              \left(
               1 - F(p_1^2,p_2^2,q^2) L(p_1^2,p_2^2)
              \right)
      \right\}
  } .
 \label{eq:form-factor}
\end{eqnarray}
 The formal limit $ g_A \rightarrow 1 $
while keeping $ M_V^2(-q^2)  $ at a fixed finite
value $ M_\rho^2 $
reduces (\ref{eq:enjl-modify}) to (\ref{triangle})
multiplied by $ L(p_1^2, p_2^2) $.
 But such a limiting procedure is not self-consistent
in the framework of the ENJL model.
 The term $ (1-g_A) $ in (\ref{eq:form-factor})
corresponds to the term necessary
in order to recover the anomalous Ward identity
missing in (\ref{triangle}) 
as claimed in Ref. \cite{Bijnens}.

 The contribution to the muon anomaly from the type of the graphs
in Fig. \ref{fig:pi0-VMD} can be written as
the magnetic moment projection
( see Sec. \ref{sec:muon-anomaly} ) of
\begin{eqnarray}
 &&
  \displaystyle{
   \frac{2}{ie}
   \times \int \mathop{\Pi}_{s=1}^{2} \frac{d^4 r_s}{(2\pi)^4}
   (ie\gamma^\lambda) \frac{i}{\ps_6 - m_\mu} (ie\gamma^\beta)
   \frac{i}{\ps_5 - m_\mu} (ie\gamma^\alpha)
  } \nonumber \\
 && \quad \quad \times
  \displaystyle{
   \frac{-i}{p_1^2} \frac{-i}{p_2^2} \frac{-i}{p_3^2}
   \frac{i}{p_4^2 - m_\pi^2}
    i A_{\alpha\beta} (p_1, p_2)
    i A_{\lambda\nu} (p_3, q)
  },
 \label{eq:enjl-a}
\end{eqnarray}
which includes the symmetry factor 2 
and an approximation $ A^2 \simeq 1 $. 
 The internal lines are labeled according to Fig. \ref{fig:pi0-VMD}.
 The terms of
$ F_{{\rm PVV}}(p_1^2, p_2^2, q^2) $ in (\ref{eq:form-factor})
consists of the point-like part, $ (1-g_A) $,
and the rest, $ g_A F(p_1^2,p_2^2,q^2) L(p_1^2,p_2^2) $.
 Thus the term proportional to $ (1-g_A)^2 $
in the product of two $ A_{\mu\nu} $'s in (\ref{eq:enjl-a})
corresponds to the contribution 
which includes two point-like vertices
which causes logarithmic divergence from the photon loop integration.
 This is handled by introducing the Feynman cutoff
for the photon propagators
\begin{equation} 
 \frac{-i}{q^2} \rightarrow \frac{-i}{q^2} - \frac{-i}{q^2 - M_c^2}
                            = \frac{-i M_c^2}{q^2 (M_c^2 - q^2)}.
\end{equation}
 This procedure
is formally the same
as that used for incorporating the vector meson dominance property
in (\ref{local-pi0}) but 
with a new mass scale $ M_c $ instead of $ M_\rho $.
 This allows us
to check the program written for the present purpose.
 When $ M_c $ is set equal to $ M_\rho $,
the result (\ref{local-pi0}) should be identical with the result
of $ g_A = 0 $,
and the result (\ref{pi0-VMD-triangle}) should correspond
to that of $ g_A = 1 $.
 This is explicitly confirmed by our program.
 We have also confirmed that the results
corresponding to various values
of $ g_A $ ( $ 0 \le g_A \le 1 $ )
always falls in the range between (\ref{pi0-VMD-triangle})
and (\ref{local-pi0}) for $ M_c = M_\rho $.
 In this way it is quite easy to observe that
for any $ g_A $ and $ M_c $, the cancelation among various terms
cannot occur for the pseudoscalar pole contribution
because all terms contribute with the same (negative) sign.
 Typical values of the $\pi^0$-pole contribution
obtained for various values of $ g_A $
and $ M_c $ are listed in Table \ref{tab:new-pole}
(10 million sampling points per iteration and 20 iterations
  for $ g_A = 0.5 $,
  2 million sampling points per iteration and 20 iterations
  for the other).
 Although our calculation is not exactly
identical with that of Ref. \cite{Bijnens}
since we have disregarded the momentum dependence
of $ g_A(-q^2) $, etc.,
our result should be approximately equal to theirs.  
 It turned out that we were not able
to reproduce the result in Ref. \cite{Bijnens}.

 The axial-vector meson contribution to four-photon vertex graph
can also be extracted from (\ref{eq:pole-sum})
\begin{eqnarray}
 &&
  \displaystyle{
   \left(
     i 4\pi\alpha \frac{g_A}{f_A}
     \bar{\Pi}^{\rm AVV}_{\alpha\mu\nu}(p_1,p_2) L(p_1^2,p_2^2)
   \right)
   \frac{-i g^{\alpha\beta}}{q^2 - M_A^2}
   \left(
     i 4\pi\alpha \frac{g_A}{f_A}
     \bar{\Pi}^{\rm AVV}_{\beta\rho\sigma}(p_3,p_4)
     L(p_3^2,p_4^2)
   \right)
  }
 \nonumber \\
 &&
  \displaystyle{
   - \left(
      i 4\pi\alpha \frac{g_A}{f_A} \frac{2m_q}{M_A}
       \left\{
         \bar{\Pi}^{\rm PVV}_{\mu\nu}(p_1,p_2) L(p_1^2,p_2^2)
         - \left.
            \bar{\Pi}^{\rm PVV}_{\mu\nu}(p_1,p_2)
           \right|_{p_1^2 = 0 = p_2^2 = q^2}
       \right\}
     \right)
  \times \frac{-i}{q^2 - M_A^2}
  }
   \nonumber \\
 && \ \ \  
  \displaystyle{
   \times
     \left(
      i 4\pi\alpha \frac{g_A}{f_A} \frac{2m_q}{M_A}
       \left\{
         \bar{\Pi}^{\rm PVV}_{\rho\sigma}(p_3,p_4) L(p_3^2,p_4^2)
         - \left.
            \bar{\Pi}^{\rm PVV}_{\rho\sigma}(p_3,p_4)
           \right|_{p_3^2 = 0 = p_4^2 = q^2}
       \right\}
     \right)
  } .
 \label{eq:AVV-muon}
\end{eqnarray}
 From Eq. (3.44) of Ref. \cite{Prades},
$ g_A/f_A $ is found to be
independent of $ G_V $.
 Thus, all contributions in Eq. (\ref{eq:AVV-muon}) vanishes
in the limit $ M_A \rightarrow \infty $ ( $ G_V \rightarrow 0 $ ),
which, of course,  should be the case.
 However the first term in Eq.(\ref{eq:AVV-muon})
may become numerically significant
since its overall coefficient
\begin{equation}
 \left( \frac{g_A}{f_A} \right)^2
 = g_A(1-g_A) \left( \frac{M_A}{f_\pi} \right)^2,
\end{equation}
where the equality is imposed by the ENJL model,
is numerically large $ \sim 46 $
(for $ g_A \sim 0.5 $ \cite{Prades} ).
 Thus there remains a possibility
that such a term contributes
to the muon $ g-2 $ with the same magnitude as pseudoscalar does,
but with the opposite sign.

 An explicit calculation shows that (for $ g_A = 0.5 $)
the first term in (\ref{eq:AVV-muon}), denoted as (1),
and the second, denoted as (2), contribute respectively as
\begin{eqnarray}
 a_\mu(a_1\ {\rm pole})[(1)] &=&
  \displaystyle{
   -0.001\ 192 \ (1) \times
    \left(
     \frac{\alpha}{\pi}
    \right)^3
  },
   \nonumber \\
 a_\mu(a_1\ {\rm pole})[(2)] &=&
  \displaystyle{
   -0.000\ 194\ (2) \times
    \left(
     \frac{\alpha}{\pi}
    \right)^3
    \ \ \ \ {\rm for}\ M_c = 1.0\ {\rm GeV}
  },
 \label{axial-pole}
\end{eqnarray}
 The first one
was calculated by 8 million sampling points per iteration
and 15 iterations,
the second by 3 million sampling points per iteration
and 15 iterations.

We find that the axial-vector contribution has the same minus sign
as the pseudoscalar one
and is one order of magnitude smaller than the latter.
 Thus such a reduction of order
as was seen in Ref. \cite{Bijnens} cannot take place 
according to our calculation.
 In this respect
we are in agreement
with the corrected result in Ref. \cite{Bijnens2}.

  The axial-vector pole contribution (\ref{axial-pole})
is negligible as a pole-type contribution,
compared to the pion pole.
 Adding the new evaluation of the $ \eta $-pole contribution 
(5 million sampling points per iteration and 20 iterations)
\begin{eqnarray}
 a_\mu(\eta\ {\rm pole}) &=&
  \displaystyle{
   - 0.011\ 69\ (1) \times \left( \frac{\alpha}{\pi} \right)^3
  }
   \nonumber \\
 &=&
 \displaystyle{
   - 14.651\ (5) \times 10^{-11} \ \ \ \ {\rm for}
                                 \ g_A = 0.5,\ M_c = 1.0\ {\rm GeV}
 },
  \label{eq:eta-pole}
\end{eqnarray}
to the $\pi^0$ pole contribution
given in Table V for $g_A = 0.5$ and $M_c =$ 1.0 GeV,
we obtain as the total pole contribution
\begin{eqnarray}
 a_\mu(b) &=&
  \displaystyle{
   - 0.045\ 9\ (91) \times \left( \frac{\alpha}{\pi} \right)^3
  } 
   \nonumber \\
 &=&
  \displaystyle{
   - 57.5\ (11.4) \times 10^{-11}
  },
 \label{eq:new-pole}
\end{eqnarray}
where the model dependence is again estimated 
to be within 20 \% of the $ M_\rho $-dependent term. 
 This replaces the value in (\ref{neutral-pole})
as the total pole contribution.

\section{Quark Loop}
\label{sec:quark-loop}

 Inferred from the ENJL model,
the quark loop diagram incorporating vector meson
can be calculated by making the substitution (\ref{eq:sub})
to photon propagators. This leads to
\begin{eqnarray}
  a_\mu(c)
   &=&
   \displaystyle{
     0.007\ 72\ (31) \left( \frac{\alpha}{\pi} \right)^3
   } 
    \nonumber \\
   &=&
    \displaystyle{
     9.68\ (39) \times 10^{-11}
    } .
  \label{q-loop-VMD}
\end{eqnarray}
To examine the quark mass dependence, 
 we define $ a_\mu(c; x m_q, M) $ with
$ a_\mu(c; m_q, M_\rho) \equiv a_\mu(c) $,
where $ m_q $ denotes the collection
of such masses
as $ m_u = m_d = 300 {\rm MeV}$
and $ m_s = 500 {\rm MeV} $
and $ x $ the common scale factor.
(Here we do not include $ c $-quark contributions
which have been included in the previous calculation \cite{K-had}
without VMD.
 Note that the $ c $-quark contribution in this case
is negligibly small
since the contribution of each quark of mass $ m_q $
is then proportional to $ m_q^{-2} $ \cite{K-had}. 
The contribution of $ c $-quark in the present model will be found
as further suppressed as is inferred from the mass dependence
presented below.)
 Numerical studies similar to the previous ones are
performed to examine quark mass dependence
by iterating integration with one billion sampling points
per iteration and 60 iterations.
 The pion mass dependence is also examined
by iterating integration with one million sampling points
per iteration and 50 iterations.

 The result is summarized in Table \ref{tab:quark-mass}
and in the following asymptotic form:
\begin{eqnarray}
 &&
  \displaystyle{
    a_\mu (c; xm_q, M_\rho) \sim
     1.94\times 10^{-2}\times x^{-4.0}
     \left ( {\alpha \over \pi} \right )^3
     \ \ \ \ {\rm for} \ \ \ \ x \geq 3
   } ,
    \label{mq-q-loop} \\
 &&
  \displaystyle{
   a_\mu(c; m_q, M) \sim
    \left[ +0.044~ 0
    - 0.43 \left ( {m_\mu \over M } \right ) \right]
    \left ( {\alpha \over \pi} \right )^3
    \ \ \ \ {\rm for} \ \ \ \ M \geq 3 M_\rho
  }.
   \label{mrho-q-loop}
\end{eqnarray}
 Note that the suppression effect of vector meson is so large here
that the value (\ref{q-loop-VMD})
is one order of magnitude smaller 
compared to (\ref{KNO-q-loop}).
 However the strong damping property on the quark mass
is consistent with the observation
that only the physical degree of freedom
is important at low energies \cite{stan}.
 Algebraically such a rapid decrease occurs
when all quark masses become comparable to $ M_\rho $
since the relevant mass scale of the system
turns then to the quark masses
so that the cancelation of the two terms in (\ref{eq:sub})
begins.

 Again, we
consider the errors arising from model-dependence
to be within 20 \% of the $ M_\rho $ dependent term.
 This is because integrations over the photon and muon momenta
are convergent in these diagrams
and hence the contribution of large photon momenta
does not distort our picture of low energy quark loop too severely.
 We are thus led to
\begin{eqnarray}
 a_\mu(c) &=&
  \displaystyle{
    \ 0.007\ 7\ (88)
             \left( \frac{\alpha}{\pi} \right)^3
  }  
   \nonumber \\
  &=&
   \displaystyle{
    9.7\ (11.1) \times 10^{-11}
   } . 
  \label{q-loop}
\end{eqnarray}
%

\section{Summary and Discussion}
\label{sec:summary}

 We have obtained the results (\ref{bestestimate}),
(\ref{neutral-pole}) and (\ref{q-loop}) as the contributions
of Fig. \ref{fig:diagrams}(a), (b) and (c), respectively.
 These diagrams have been discussed
in Sec. \ref{sec:review} to contribute most significantly
and independently to the hadronic light-by-light scattering effect
on muon anomaly $ a_\mu $,
as guided by the use of chiral and $ 1/N_c $ expansion.
The $ M_\rho $ dependence of the contributions
(\ref{approx}), (\ref{mrho-for-pi0}) and (\ref{mrho-q-loop}) 
indicates that
the integration over the photon momenta
receives considerable contribution
from the region where photons are far off shell.
 We have estimated that these high mass contributions should be
well within 20\% of the vector meson contribution,
which leads to the large uncertainties assigned to 
(\ref{bestestimate}), (\ref{eq:new-pole})
and (\ref{q-loop}).
 Combining these results we obtain
\begin{equation}
 a_\mu({\rm light\mbox{-}by\mbox{-}light}) =
  - 52\ (18) \times 10^{-11} .
 \label{total-result}
\end{equation}
 This is almost within the error (\ref{expectederror})
in the upcoming experiment.
 Therefore, with the progress of measurement of $ R $
\cite{worstell},
the accurate determination of muon anomaly by future experiment
will actually show the presence of
the weak interaction correction \cite{weak,weak2} 
and  serves as a new constraint on physics beyond the standard 
model.

 Let us now discuss possible causes of difference
between our result and the recent result
of Bijnens $et~al$. \cite{Bijnens2}, which is based on the ENJL model.
 For comparison's sake,
let us list their results corresponding to Figs. 2(a), 2(b), and 2(c):

\begin{equation}
 a_\mu(a)_{BPP} = (-14.5 \sim - 22.8 ) \times 10^{-11} 
\label{BPPa}
\end{equation}
for the cut-off $\mu$ ranging from 0.6 GeV to 4.0 GeV, and

\begin{equation}
 a_\mu(b)_{BPP} = (-72 \sim - 186 ) \times 10^{-11} 
\label{BPPb}
\end{equation}
and
\begin{equation}
 a_\mu(c)_{BPP} = (11.4 \sim  20.0 ) \times 10^{-11} 
\label{BPPc}
\end{equation}
for the cut-off $\mu$ ranging from 0.7 GeV to 8.0 GeV.

 On the surface, the results of \cite{Bijnens2} seems
to be more reliable than ours,
being less dependent on assumptions outside of the ENJL model. 
 On the other hand, their result is not free
from ambiguities either mainly because their theory does not tell
which cut-off should be favored.  
 In particular, it seems to be difficult
to justify their results for large $\mu$
which lies well beyond the region of applicability of the ENJL model.

 The result (\ref{BPPa}) is about 4 times larger
than our result (\ref{bestestimate}).
 This may partly be due to our simplifying assumptions,
such as the complete vector meson dominance ($a$ = 2)
and the neglect of momentum dependence of various masses
and effective coupling constants. 
 Note, however, that the Lagrangian of \cite{Bijnens2}
seems to have the $\pi^+ \pi^- \rho^0 \rho^0$ vertex.  
 The presence of such a coupling (without derivatives)
will be inconsistent with the low energy phenomenology. 
 It also means that their Lagrangian
does not satisfy the Ward identity (\ref{Ward-identity})
contrary to their assertion.
 In particular, their Lagrangian does not seem to incorporate
the vector meson consistently,
as is described in detail in Appendix A.
 If this is the case, it could explain the bulk of the difference.
 It should also be recalled that the smallness of our result
(\ref{bestestimate}) is a consequence of an accidental cancelation
of two main terms for the physical $\rho$ mass value.  
Such a delicate cancelation is not visible
in the calculation of \cite{Bijnens2}.

 The contribution (\ref{BPPb}) is 2 to 5 times larger
than our estimate (\ref{neutral-pole}). 
 Since this is the largest term,
it is the main source of disagreement between the two calculations.
 Actually, the low end value  ($-72 \times 10^{-11}$) of (\ref{BPPb})
is of the same order of magnitude as our value for 
$a_\mu(b; m_\pi,\infty, m_q)$
given in (\ref{twoanomalyloop}). 
 Recall that in the latter calculation
the anomalous $\pi^0 \gamma^* \gamma^*$ vertex
is approximated by
a triangular loop of constituent quarks
and photons are attached to the $``$bare$''$ quark directly.
 If one assumes that any QCD modification
softens this coupling,
our result (\ref{twoanomalyloop}) may be regarded
as some sort of upper limit of the contribution of Fig. 2(b). 
 On the other hand, the result (\ref{BPPb})
increases with increasing cut-off $\mu$ beyond this $``$bound",
 suggesting that the result (\ref{BPPb})
diverges logarithmically as $\mu \rightarrow \infty$.
This behavior is a consequence of the presence of 
a hard PVV vertex in their Lagrangian.
Its prediction on the muon $g - 2$ must be viewed
with severe reservation,
however, since it is obtained by applying
the ENJL model beyond its domain of validity determined
by the cut-off $\Lambda_{\chi}$.
 In fact, such an unwarranted application of the model
(with a hard anomaly term) violates unitarity as $\gamma^*$
goes far off shell \cite{hikasa},
and hence must be tempered with some form factor.  
 In other words, any realistic theory must
be consistent with unitarity,
be it the ENJL model or the HLS model.

 An examination of Fig. 3,
in the limit where both fermion triangles shrink to points, 
shows that the UV divergence arises from the integration domain
in which the 
momenta carried by the photon 3 and pion 4
are small
while the momenta carried by the photons 1 and 2 are large.
 The far-off-shell structure
of the $\pi^0 \gamma^* \gamma^*$ vertex
in such a region has been studied
using the Bjorken-Johnson-Low theorem \cite{gerard},
which shows $1/q^2$ behavior
asymptotically, where $q \sim q_1 \sim q_2$ \cite{marciano}.
 The case where only one of the photons ($q_2$)
is far off-shell has also been studied
by an operator product expansion technique \cite{Brodsky}.
 Based on the latter analysis a formula 
of the form
interpolating between $p_1^2 = 0$ and $p_1^2 = \infty$ 
\begin{eqnarray}
   &&
  \displaystyle{
    F(p_1^2 \rightarrow \infty , p_2^2 = 0, q^2) =
    {1 \over {1 - ( p_1^2 /(8\pi^2 f_\pi^2))}}
  }, \nonumber \\
   &&
  \displaystyle{
   ~~~~~~~~~~~~~~~~~~~~~~~~~~ \sim { 1 \over {1 - ( p_1^2 /M^2 )}}
  },
\label{asymp-form}
\end{eqnarray}
has been suggested   
for the form factor $ F(p_1^2, p_2^2, q^2) $
normalized similarly as that of (\ref{PVV-form}).
 The experimental data fit Eq. (\ref{asymp-form}) very well
with $M^2 \sim (0.77\,{\rm GeV}/c)^2$
over the range $2.0 ({\rm GeV}/c)^2$
to $20.0 ({\rm GeV}/c)^2$ \cite{experiment}.
 In Ref. \cite{Harada} it is argued that 
the off-shell behavior
of the $\pi^0 \gamma^* \gamma^*$ amplitude  is represented
reasonably well by the quark triangle amplitude (\ref{triangle}) 
if one takes account of the asymptotic freedom of QCD
and a nonperturbative generation of constituent quark mass. 
 The result of their analysis is consistent with those quoted above.  
 These considerations
suggest that our model based on Eq. (\ref{triangle})
may in fact be a reasonably good representation
of the contribution of Fig. 2(b) \cite{Asymptotic}. 

 There is relatively small difference
between (\ref{BPPc}) and (\ref{q-loop}).
 The remaining difference is within the range of uncertainty
caused by our simplifying assumptions.  
 In fact the good agreement between (\ref{q-loop}) and (\ref{BPPc})
may even be an indication that we have overestimated
the model dependence in (\ref{q-loop}).
 As was mentioned already,
this is consistent with the fact
that integrations over the photon
and muon momenta are convergent
and do not distort low energy quark-loop picture too severely. 

 It appears to be difficult to resolve
the difference between our calculation
and that of Ref. \cite{Bijnens2} completely
because of different approaches
and because of the necessity
to apply the low energy effective theory of strong interaction
beyond its safely tested domain.
 The complete resolution may have
to wait for the lattice QCD calculation of the four-point function.
 With the rapid improvement of the computing power,
such a day may not be too far off.

\acknowledgements

 We acknowledge useful discussions with Bijnens, Pallante, and Prades.
 T. K. thanks S. Tanabashi and H. Kawai for helpful discussions. 
 T. K.'s work is supported
in part by the U. S. National Science Foundation. 
 Part of numerical computation was carried out
at the Cornell National Supercomputing Facility,
which receives major funding
from the U. S. National Science Foundation and the IBM Corporation,
with additional support
from New York State and members of the Corporate Research Institute. 
 M. H. is supported in part
by Japan Society for the Promotion of Science
for Japanese Junior Scientists.
 He thanks the Toyota Physical and Chemical Research 
Institute for the support he has received in the early stage 
of this work.
 A. I. S. acknowledges a partial support from Daiko Foundation.

\appendix

\section{Vector meson in the BPP Chiral Lagrangian}

 In this Appendix we show that
the vector meson is not properly incorporated as a dynamical field in 
the chiral Lagrangian (5.6) of Ref. \cite{Bijnens2}
\begin{eqnarray}
 {\cal L} &=&
  \frac{f_\pi^2}{4}
  {\rm tr}
  \left[
   {\cal D}_\mu U {\cal D}^\mu U^\dagger + {\cal M} U^\dagger
   + U {\cal M}^\dagger
  \right]
  \nonumber \\
  && + \frac{M_V^2}{2}
    \left[
     \left(
      \rho_\mu - \frac{1}{g} v_\mu
     \right)
     \left(
      \rho^\mu - \frac{1}{g} v^\mu
     \right)
    \right]
    + \frac{1}{8g^2}
      {\rm tr}
      \left[
       L_{\mu\nu} L^{\mu\nu} + R_{\mu\nu} R^{\mu\nu}
      \right] .
 \label{B-chiral}
\end{eqnarray}
 Here $\rho_\mu$ denotes vector meson field
and $ {\cal M} $ represents the quark mass matrix,
$ {\cal M}={\rm diag}(m_u,m_d,m_s) $, in the three-flavor case.
 $v_\mu$ and $a_\mu$ are the external vector and axial-vector fields
respectively, and $ L_{\mu\nu} $ and $ R_{\mu\nu} $ are given by

\begin{eqnarray}
 L_{\mu\nu} &=&
  \del_\mu(g\rho_\nu-a_\nu) - \del_\nu(g\rho_\mu-a_\mu)
  + [(g\rho_\mu-a_\mu),\ (g\rho_\nu-a_\nu)]
  \nonumber \\
 R_{\mu\nu} &=&
  \del_\mu(g\rho_\nu+a_\nu) - \del_\nu(g\rho_\mu+a_\mu)
  + [(g\rho_\mu+a_\mu),\ (g\rho_\nu+a_\nu)] .
\end{eqnarray}

 The transformation properties of $v_\mu$, $a_\mu$, $\rho_\mu$,
and the unitary matrix ($U$) consisting of the pseudoscalar meson
are given by
\begin{eqnarray}
 && U \rightarrow  U^\prime = V_R U V_L^\dagger , \\
 &&(v_\mu + a_\mu)
     \rightarrow 
     (v^\prime_\mu + a^\prime_\mu) =
     V_R (v_\mu + a_\mu) V_R^\dagger + i V_R \partial_\mu V_R^\dagger ,
   \label{v+} \\
 &&(v_\mu - a_\mu)
     \rightarrow
     (v^\prime_\mu - a^\prime_\mu) =
     V_L (v_\mu - a_\mu) V_L^\dagger + i V_L \partial_\mu V_L^\dagger ,
    \label{v-} \\
 &&(g \rho_\mu + a_\mu)
     \rightarrow
     (g \rho^\prime_\mu + a^\prime_\mu) =
     V_R (g \rho_\mu + a_\mu) V_R^\dagger
             + i V_R \partial_\mu V_R^\dagger ,
    \label{rho+}\\
 &&(g \rho_\mu - a_\mu)
      \rightarrow
     (a \rho^\prime_\mu - a^\prime_\mu) =
      V_L (g \rho_\mu - a_\mu) V_L^\dagger
             + i V_L \partial_\mu V_L^\dagger
    \label{rho-} 
\end{eqnarray}
for the chiral transformation ($V_L, V_R$).
 Then the covariant derivative ${\cal D}_\mu U$:
\begin{equation}
 {\cal D}_\mu U \equiv \del_\mu U - i (g\rho_\mu + a_\mu)U
                             + i U(g\rho_\mu - a_\mu),
\end{equation}
in which vector meson $\rho_\mu$ appears instead of $v_\mu$
to realize the vector meson dominance,
transforms covariantly.
 The vector meson mass term in (\ref{B-chiral})
is also chiral-invariant as shown below.
(As a matter of fact it vanishes.)

 The transformation property of the combination $(g\rho_\mu - v_\mu)$
is found to be 
\begin{eqnarray}
 (g\rho_\mu^\prime - v_\mu^\prime)
 &=& (g\rho_\mu^\prime + a_\mu^\prime) - (v_\mu^\prime + a_\mu^\prime)
  \nonumber \\
 &=& V_R (g\rho_\mu - a_\mu)V_R^\dagger 
  \label{RR}
\end{eqnarray}
from (\ref{v+}) and (\ref{rho+}). 
 On the other hand, from (\ref{v-}) and (\ref{rho-}), we find
\begin{eqnarray}
 (g\rho_\mu^\prime - v_\mu^\prime)
 &=& (g\rho_\mu^\prime - a_\mu^\prime) - (v_\mu^\prime - a_\mu^\prime)
  \nonumber \\
 &=& V_L (g\rho_\mu - a_\mu)V_L^\dagger .
  \label{LL}
\end{eqnarray}
 Since $V_L$ and $V_R$ are independent,
we may consider the case where $V_L = 1$ and $V_R$ is nontrivial.
 Then, from Eqs. (\ref{RR}) and (\ref{LL}) we obtain
\begin{equation}
 (g \rho_\mu - v_\mu)
  = V_R (g \rho_\mu - v_\mu) V_R^\dagger .
  \label{VR}
\end{equation}
 For simplicity, let us consider the two-flavor case,
and set $ V_R = e^{i \frac{\pi}{2} \sigma^2} $
(isospin rotation about the second axis).
 Then the RHS of eq. (\ref{VR}) becomes
\begin{equation}
  - (g\rho_\mu - v_\mu))^a\ 
    \frac{\sigma^a}{2}^* ,
\end{equation}
where ``$ ^* $'' denotes complex conjugation. 

 Picking up the third component of
both sides of (\ref{VR}), for instance, we get
\begin{equation}
 (g\rho_\mu - v_\mu)^3 = - (g\rho_\mu - v_\mu)^3,
\end{equation}
that is, $ (g\rho_\mu - v_\mu)^3 = 0 $.

 In a similar way we can prove that $ g\rho_\mu - v_\mu $
vanishes for other components.
 This means that $\rho_\mu$ is nothing but
an external vector field and
the vector meson has not been incorporated
in the theory as a dynamical object.

%


\begin{table}
 \caption{ Orders with respect to $ 1/N_c $ and chiral expansions
           of the diagrams
           shown in Fig. {\protect\ref{fig:diagrams}}.
         }
 \label{tab:orders}
 \vspace{0.3cm}
 \begin{tabular}{lcc}
   Diagram & $ 1/N_c $ expansion & Chiral expansion \\
   \hline
   Fig. \ref{fig:diagrams}(a) & 1 & $ p^4 $ \\
   \hline
   Fig. \ref{fig:diagrams}(b) & $ N_c $ & $ p^6 $ \\
   \hline
   Fig. \ref{fig:diagrams}(c) & $ N_c $ & $ p^8 $
 \end{tabular}
\end{table}

\begin{table}
 \caption{ $ m_\pi $ and $ M_\rho $
          dependence of $ \pi^\pm $ loop contribution.
          Table lists 
          $ (x m_\pi/m_\mu)^2 $
          $ \times a_\mu(a; xm_\pi, M) $
          for $ M = M_\rho $ and $ M = \infty $,
          and
          $ (x M_\rho/m_\mu) $
          $ \times
            [a_\mu(a; m_\pi, x M_\rho) - a_\mu(a; m_\pi, \infty)] $.}
 \label{tab:pi-mass}
 \begin{tabular}{cccc}
  $ x $ &
  \begin{tabular}{l}
   $ (\pi/\alpha)^3 \times (x m_\pi/m_\mu)^2 $ \\
   \  $ \times a_\mu(a; xm_\pi, M_\rho) $
  \end{tabular}
  &
  \begin{tabular}{l}
   $ (\pi/\alpha)^3 \times (x m_\pi/m_\mu)^2 $ \\
   \ $ \times a_\mu(a; xm_\pi, \infty) $
  \end{tabular}
  &  
  \begin{tabular}{l}
   $ (\pi/\alpha)^3 \times (x M_\rho/m_\mu) $ \\
   \ $ \times [a_\mu(a; m_\pi, x M_\rho) - a_\mu(a; m_\pi, \infty)] $
  \end{tabular}
  \\
  \hline
   5 & 0.064~9\ (44) & $-$0.082\ (8) & 0.234\ (7) \\
  \hline
  10 & 0.094~5\ (45) & $-$0.093\ (9) & 0.190\ (7) \\
  \hline
  15 & 0.105\ (4) & $-$0.099\ (14) & 0.155\ (6) \\
  \hline
  20 & 0.106\ (5) & $-$0.094\ (17) & 0.141\ (6) \\
  \hline
  25 & 0.107\ (5) & $-$0.094\ (18) & 0.124\ (6) \\
  \hline
  30 & 0.110\ (5) & $-$0.094\ (25) & 0.120\ (6) \\
 \end{tabular}
\end{table}

\begin{table}
 \caption{ $ m_\pi $ and $ M_\rho $
          dependence of $ \pi^0 $ pole contribution.
           Table lists
          $ (x m_\pi/m_\mu)^2 $
          $ \times a_\mu(b; xm_\pi, M_\rho, m_q) $ and
          $ (x M_\rho/m_\mu) $
          $ \times
            [a_\mu(b; m_\pi, x M_\rho, m_q) $
          $ - a_\mu(b; m_\pi, \infty, m_q)] $ . }
 \label{tab:pi0-mass}
 \begin{tabular}{ccc}
  $ x $ &
  \begin{tabular}{l}
   $ \left( \pi/\alpha \right)^3 \cdot (x m_\pi/m_\mu)^2 \times $ \\
   $ a_\mu(b; xm_\pi, M_\rho, m_q) $
  \end{tabular} &
  \begin{tabular}{l}
   $ \left( \pi/\alpha \right)^3 \cdot (x M_\rho/m_\mu) \times $ \\
   $ [a_\mu(b; m_\pi, x M_\rho, m_q)
       - a_\mu(b; m_\pi, \infty, m_q)] $
  \end{tabular}
  \\
  \hline
   5 & $-$0.168~4\ (1) & 0.336~7\ (2)
  \\
  \hline
  10 & $-$0.203~4\ (1) & 0.276~5\ (2)
  \\
  \hline
  15 & $-$0.213~8\ (1) & 0.235~9\ (2)
  \\
  \hline
  20 & $-$0.218~2\ (2) & 0.207~3\ (2)
  \\
  \hline
  25 & $-$0.220~2\ (2) & 0.186~7\ (2)
  \\
  \hline
  30 & $-$0.221~5\ (2) & 0.170~1\ (2)
  \\
 \end{tabular}
\end{table}

\begin{table}
 \caption{ $ M_\rho $
           dependence of $ \eta $ pole contribution.
           Table lists
          $ (x M_\rho/m_\mu) $
          $ \times
            [a_\mu(b; m_\pi, x M_\rho, m_q) $
          $ - a_\mu(b; m_\pi, \infty, m_q)] $ . }
 \label{tab:eta}
 \begin{tabular}{cc}
  $ x $ &
  $ \left( \pi/\alpha \right)^3 \cdot (x M_\rho/m_\mu) \times
    [a_\mu(b; m_\pi, x M_\rho, m_q)
      - a_\mu(b; m_\pi, \infty, m_q)] $
  \\
  \hline
   5 & 0.135\ 7\ (8)
  \\
  \hline
  10 & 0.114\ 5\ (8)
  \\
  \hline
  15 & 0.108\ 6\ (7)
  \\
  \hline
  20 & 0.087\ 3\ (5)
  \\
  \hline
  25 & 0.078\ 9\ (5)
  \\
  \hline
  30 & 0.072\ 0\ (5)
  \\
 \end{tabular}
\end{table}

\begin{table}
 \caption{ $ \pi^0 $-pole contribution for various values
           of $ g_A $ and $ M_c $.
           All the values are listed with the factor $ (\alpha/\pi)^3 $
           removed. }
 \label{tab:new-pole}
 \begin{tabular}{cccc}
  $ g_A $ & $ M_c = 1 $ GeV & $ M_c = 2 $ GeV & $ M_c = 4 $ GeV
  \\
  \hline
  0.3 & $-$ 0.053\ 85\ (2) & $-$ 0.089\ 58\ (3) & $-$ 0.137\ 35\ (4)
  \\
  \hline
  0.5 & $-$ 0.034\ 18\ (1) & $-$ 0.057\ 06\ (2) & $-$ 0.086\ 20\ (2)
  \\
  \hline
  0.8 & $-$ 0.036\ 43\ (3) & $-$ 0.044\ 03\ (3) & $-$ 0.052\ 67\ (3)
  \\
 \end{tabular}
\end{table}

\begin{table}
 \caption{ Quark mass and $ M_\rho $
          dependence of quark loop contribution.
           Table lists
          $ x^4 $
          $ \times a_\mu(c; xm_q, M_\rho) $
          and
          $ (x M_\rho/m_\mu) $
          $ \times
            [a_\mu(c; m_q, x M_\rho) - a_\mu(c; m_q, \infty)] $.}
 \label{tab:quark-mass}
 \begin{tabular}{ccc}
  $ x $ &
  \begin{tabular}{r}
   $ ( \pi/\alpha)^3 \times
     x^4 a_\mu(c; x m_q, M_\rho) $ \\
   \quad $ \times 10 $
  \end{tabular}
  &
  \begin{tabular}{l}
   $ ( \pi/\alpha)^3 \times (x M_\rho/m_\mu) $ \\
   $ \times [a_\mu(c; m_q, x M_\rho) - a_\mu(c; m_q, \infty)] $
  \end{tabular}
  \\
  \hline
   5 & 0.177\ (6) & $-$ 0.489\ (5) \\
  \hline
  10 & 0.198\ (23) & $-$ 0.486\ (5) \\
  \hline
  15 & 0.191\ (51) & $-$ 0.460\ (5) \\
  \hline
  20 & 0.193\ (91) & $-$ 0.423\ (4) \\
  \hline
  25 & 0.195\ (142) & $-$ 0.402\ (4) \\
  \hline
  30 & 0.194\ (205) & $-$ 0.385\ (4) \\
 \end{tabular}
\end{table}

%
%
%
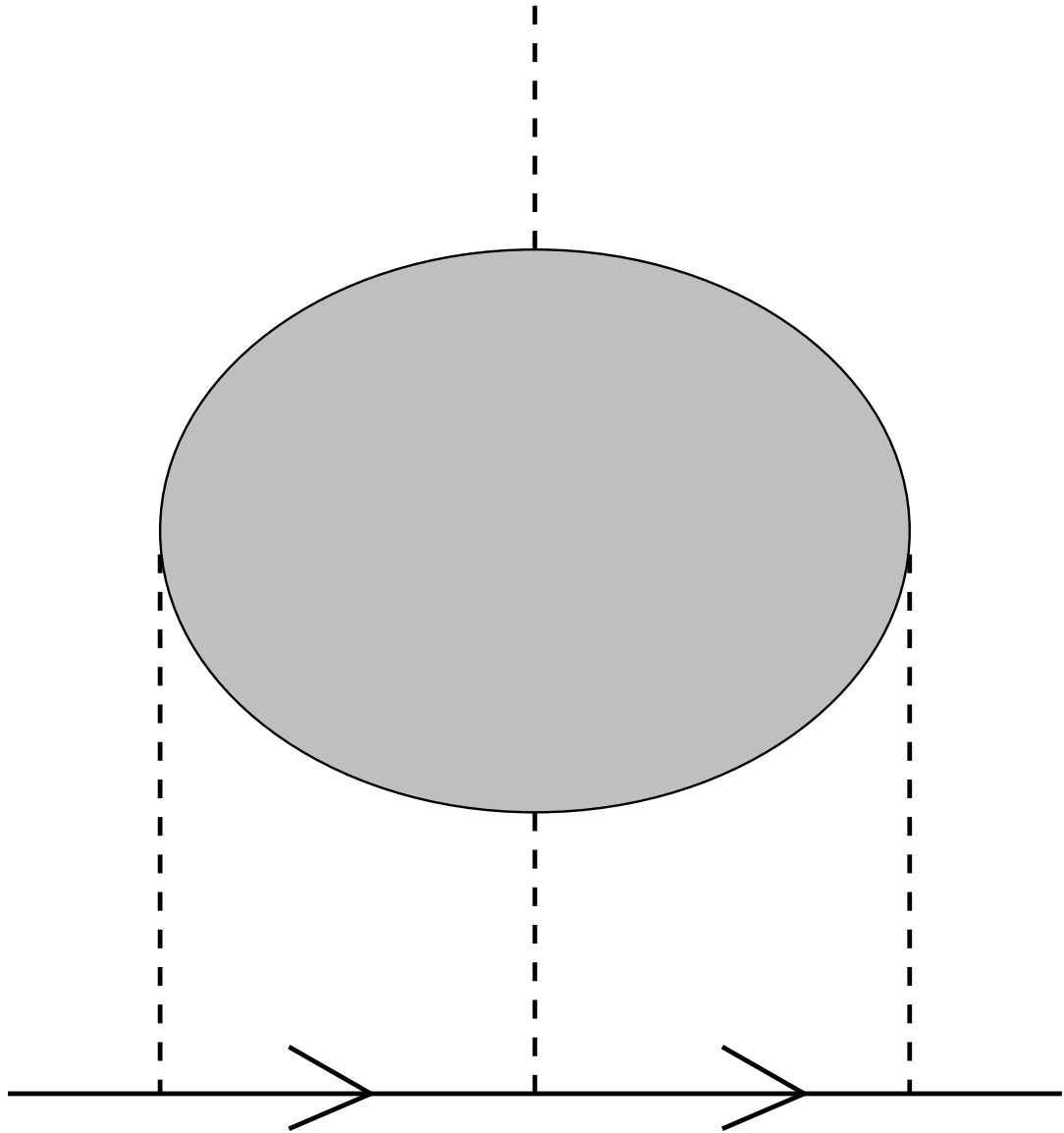
\begin{figure}[hbp]
  \caption{ Hadronic light-by-light scattering
            (shown by the shaded blob) contribution
            to the muon anomaly.
            Solid line and dashed line represent
            muon and photon, respectively. }
  \label{fig:light}
\end{figure}
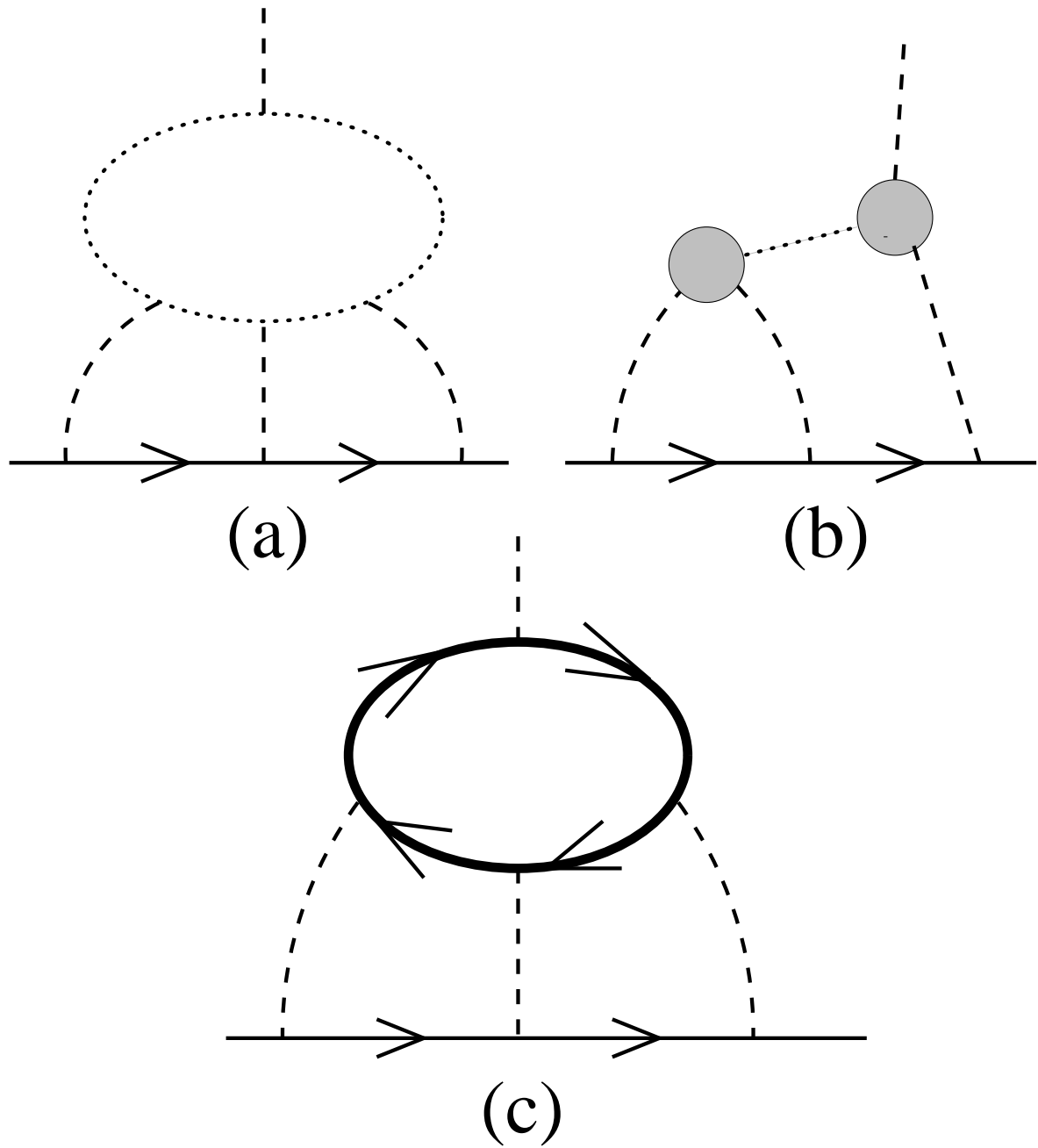
\begin{figure}[hbp]
 \caption{
 Representative diagrams which 
 dominate the hadronic light-by-light effect on $ a_\mu $
 at low energies.
 Other diagrams are obtained by permutation of the photon legs. 
 (a) Charged pseudoscalar diagram 
 in which the dotted line corresponds to $ \pi^\pm $, etc.
 (b) One of the $ \pi^0 $ pole graphs,
 in which the dotted line corresponds to $ \pi^0 $
 and the blob represents the $ \pi \gamma \gamma $ vertex.
 (c) Quark loop contribution,
 where quark is denoted by bold line.}
 \label{fig:diagrams}
\end{figure}

\begin{figure}[hbp]
  \caption{ Diagram of neutral pseudoscalar pole contribution
            with VMD and quark triangular loop.
            The bold dashed line represents the vector meson.
            The arrow attached to each internal line label
            indicates the direction of the corresponding momentum. 
            Other diagrams are obtained
            by permutation of the photon legs. }
  \label{fig:pi0-VMD}
\end{figure}

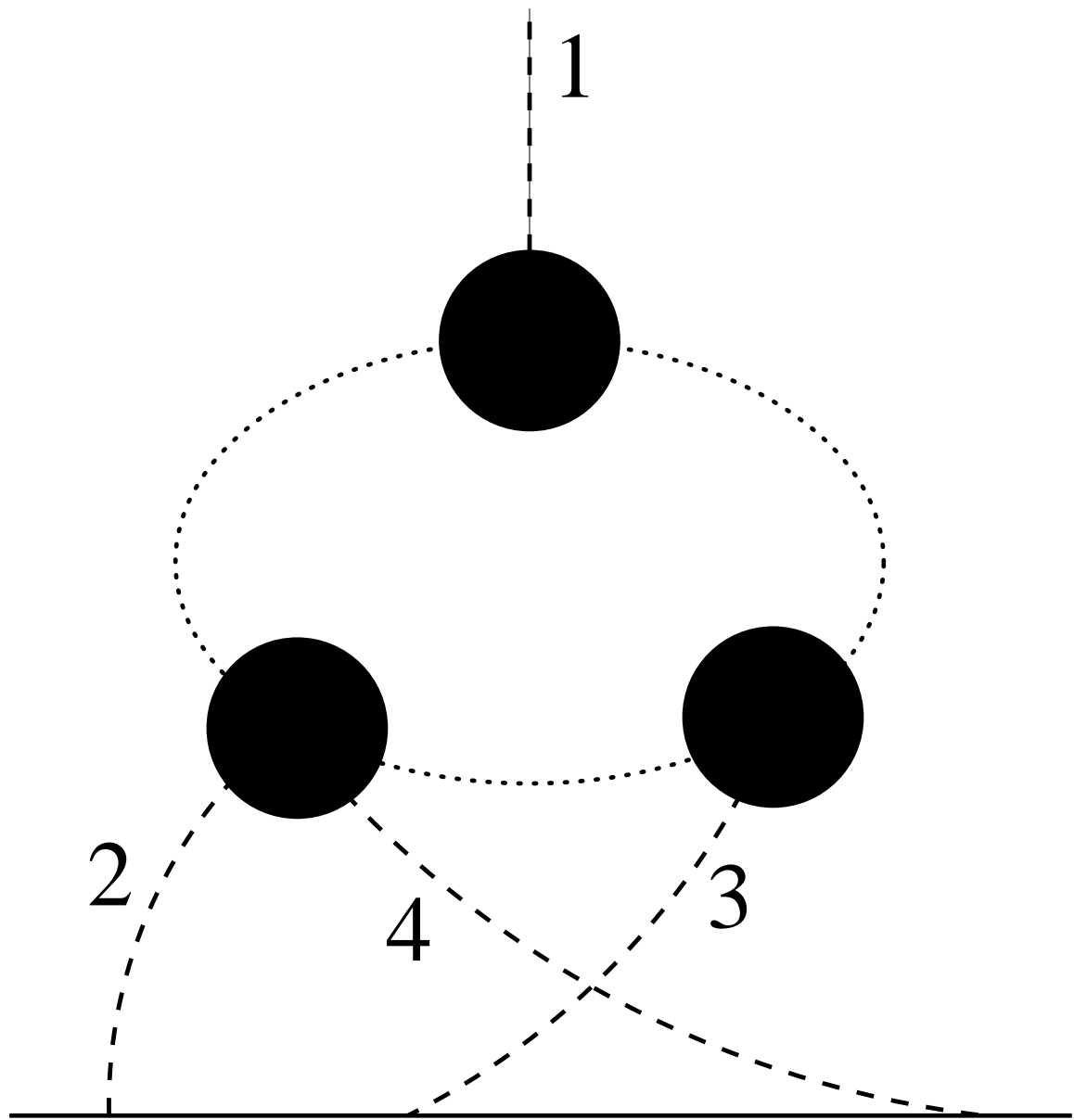
\begin{figure}[hbp]
  \caption{ A typical diagram contributing to $ a_\mu(a) $.
            To facilitate correspondence with the text,
            a number is attached to each internal line. 
            Other diagrams are obtained by permutation
            of the photon legs. }
  \label{fig:A2}
\end{figure}

\vfill
\newpage
M. H.
[There is no meaning in the above letters.
It is inserted for printing out all pages in the two-sides.
( The total number of the files have to be odd, then.
Sorry.]


\begin{references}
\bibitem{gauge-choice}
  Feynman gauge is chosen for the electromagnetic gauge fixing.
  In the calculation of the muon anomaly, the $ k_\mu k_\nu $ part of
  the photon propagator is irrelevant due to the Ward identity.
  Likewise the $ k_\mu k_\nu $ part of the $ \rho $ meson propagator
  drops out automatically in that case.
\bibitem{foot}
   We also expect, as in the ENJL model,  
  that a $q^2$-dependent function $M_V^2 (q^2)$,
  where $q$ is the $\rho$ momentum,
  appears in place of the constant $\rho$ meson mass.
   As is seen from the explicit form of $M_V^2 (q^2 )$
  given in \cite{deRafael},
  the $q^2$-dependence in the relevant range 
  ($ \leq M_\rho$) is relatively mild
  so that the correction to this approximation seems to be small.
\bibitem{Asymptotic}
  The asymptotic behaviors given in \cite{gerard} and \cite{Brodsky}
  differ by some constant factor indicating
  the presence of some unresolved problem. 
  However, both show $q^{-2}$-type damping for large $q^2$,
  and that is the crucial point as far as our discussion is concerned.
\end{references}
\end{document}